\documentclass[a4paper,11pt]{article}
\pdfoutput=1 

\usepackage{jheppub} 

\usepackage[T1]{fontenc} 

\usepackage{epsfig}
\usepackage[normalem]{ulem}
\usepackage{tikz}
\usetikzlibrary{arrows}
\usetikzlibrary{decorations.markings}
\usepackage{multirow}
\usepackage{makecell}

\newcommand{\be}{\begin{equation}}
\newcommand{\ee}{\end{equation}}
\newcommand{\nn}{\nonumber}

\newcommand{\of}{{\rm d}}
\renewcommand{\d}[2][]{\mathrm{d}^{#1}{#2}}

\title{\boldmath A New Spin on the Weak Gravity Conjecture}

\author[a]{Lars Aalsma,}
\author[b]{Alex Cole,}
\author[a]{Gregory J. Loges,}
\author[a]{Gary Shiu}

\affiliation[a]{Department of Physics, University of Wisconsin, 1150 University Ave, Madison WI 53706, U.S.A.}
\affiliation[b]{GRAPPA and ITFA, Institute of Physics, University of Amsterdam, Science Park 904, 1090 GL Amsterdam, the Netherlands}

\emailAdd{laalsma@wisc.edu}
\emailAdd{a.e.cole@uva.nl}
\emailAdd{gloges@wisc.edu}
\emailAdd{shiu@physics.wisc.edu}

\abstract{The mild form of the Weak Gravity Conjecture states that quantum or higher-derivative corrections should decrease the mass of large extremal charged black holes at fixed charge. This allows extremal black holes to decay, unless protected by a symmetry (such as supersymmetry). We reformulate this conjecture as an integrated condition on the effective stress tensor capturing the effect of quantum or higher-derivative corrections. In addition to charged black holes, we also consider rotating BTZ black holes and show that this condition is satisfied as a consequence of the $c$-theorem, proving a spinning version of the Weak Gravity Conjecture. We also apply our results to a five-dimensional boosted black string with higher-derivative corrections. The boosted black string has a BTZ$\times S^2$ near-horizon geometry and, after Kaluza-Klein reduction, describes a four-dimensional charged black hole. Combining the spinning and charged Weak Gravity Conjecture we obtain positivity bounds on the five-dimensional Wilson coefficients that are stronger than those obtained from charged black holes alone.}

\begin{document}

\maketitle
\flushbottom

\section{Introduction}
\label{sec:intro}
The effective field theory (EFT) approach allows one to systematically compute UV effects on the IR dynamics of a physical system. In an EFT, UV effects are encoded in an infinite series of higher-dimensional operators and corresponding Wilson coefficients. At first sight, it seems that without performing any measurements a low-energy observer cannot know the value of any Wilson coefficients. However, it has been known for a long time that not every EFT is ``healthy'' in the sense that it enjoys an embedding in a UV-complete theory free of pathologies. For example, unitarity and causality can constrain certain (combinations of) Wilson coefficients to be positive \cite{Adams:2006sv}. In the context of quantum gravity the criteria that distinguish healthy EFTs from sick ones are known as swampland conjectures (see \cite{Brennan:2017rbf,Palti:2019pca} for reviews). Healthy EFTs that enjoy an embedding in a consistent theory of quantum gravity are said to reside in the landscape, while EFTs that cannot be embedded in quantum gravity belong to the swampland. In the absence of experimental data sufficiently sensitive to directly probe quantum gravity, swampland criteria are helpful in constraining the space of EFTs that arise in its low-energy limits.

The swampland conjecture that is the focus of this paper is the Weak Gravity Conjecture (WGC) \cite{ArkaniHamed:2006dz}, which in its original form states that any theory with a $U(1)$ gauge field must include at least one state whose charge-to-mass ratio exceeds that of extremal black holes in that theory. This allows extremal black holes to decay, unless protected by a symmetry (such as supersymmetry). Further refinements of the WGC specify the energy scales at which these states should appear. Strong forms of the conjecture require the states in question to be light or part of a tower \cite{Andriolo:2018lvp} or charge sublattice \cite{Heidenreich:2016aqi,Montero:2016tif} of (super)extremal states. Milder forms of the WGC allow the states to be heavy or even given by black holes with an extremality bound that is corrected by quantum or higher-derivative corrections.\footnote{These milder forms of the WGC can in some cases be upgraded to stronger forms using modular invariance and the matching of anomalies \cite{Aalsma:2019ryi}.} This latter version is referred to as the ``mild form'' of the WGC and requires that corrections increase the charge-to-mass ratio of extremal black holes in a canonical ensemble (fixed charge and temperature). Because the sign of the corrections to the extremality bound depends on the sign of the Wilson coefficients, unitarity and causality play a crucial role.

In fact, several checks \cite{Kats:2006xp,Cano:2019oma,Cano:2019ycn} and proofs applying in different restricted settings and making use of thermodynamics \cite{Cheung:2018cwt,Loges:2019jzs} or unitarity and causality \cite{Hamada:2018dde,Bellazzini:2019xts,Loges:2019jzs} have been given by now, but it has become clear that generically one needs additional UV information and that the WGC cannot follow solely from IR consistency. In the presence of a massless graviton, positivity
bounds cannot completely constrain the correction to the extremality bound due to a singularity in the forward limit of graviton exchange in the t-channel (see \cite{Loges:2019jzs,Hamada:2018dde,Bellazzini:2019xts,Alberte:2020jsk,Tokuda:2020mlf} for recent discussions).\footnote{In \cite{Loges:2019jzs,Hamada:2018dde} an assumption about the UV-theory completing the higher-derivative corrected theory was made and in \cite{Andriolo:2020lul,Loges:2020trf} the effective action was imposed to be duality invariant.}

\begin{figure}[t]
\centering
\includegraphics[scale=.9]{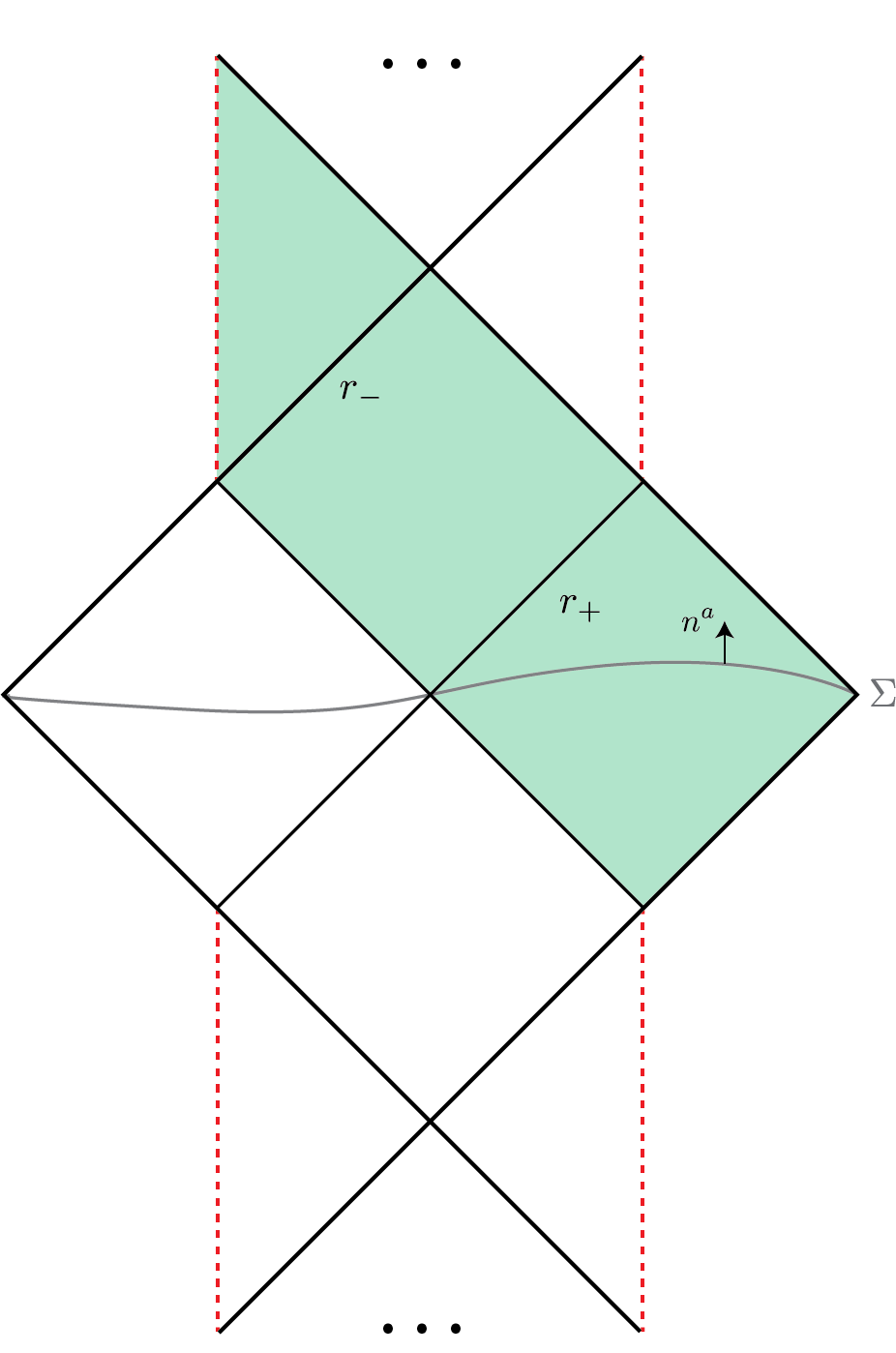}
\caption{Penrose diagram of the maximally extended Reissner-Nordstr\"om geometry. Its timelike singularities are red dotted lines, the inner and outer horizon given by $r_\pm$, and the gray surface $\Sigma$ is a Cauchy slice with normal vector $n^a$. In this paper we restrict to the green shaded region. Classically, the Penrose diagram extends indefinitely to the past and future.}
\label{fig:RN-Penrose}
\end{figure}

It is thus of interest to identify the minimal set of assumptions needed to prove the WGC. To manage expectations, we will not identify this minimal set of assumptions in this paper. Instead, we will reinterpret the mild form of the WGC as a criterion on matter that generates corrections to the extremality bound. In a way, this is similar to using energy conditions to exclude pathological matter contributions (see \cite{Loges:2020trf} for example). This results in a condition on the stress tensor that is equivalent to the WGC. For a $d$-dimensional black hole this condition is given by
\be \label{eq:WGCcondition}
\int_\Sigma \mathrm{d}^{d-1}x \,\sqrt{h}\,\delta T^{\rm eff}_{ab}\xi^an^b \leq 0 ~.
\ee
Here $\Sigma$ is a Cauchy slice with normal vector $n^a$ and $\xi^a$ is a Killing vector for which the horizon is a Killing horizon, see Fig.\ \ref{fig:RN-Penrose}.
$\delta T^{\rm eff}_{ab}$ is an effective stress tensor whose definition will be given in the main body of this article. This condition has several attractive features. Just as in the thermodynamic approach described in \cite{Reall:2019sah} it is only neccesary to know the uncorrected black hole metric to derive corrections to the extremality bound. This has the technical advantage that one does not need to solve the (possibly complicated) corrected Einstein equations in order to evaluate the WGC. In addition, this condition is valid for any correction that generates an effective stress tensor, not just higher-derivative corrections, so it can be applied to a wide range of scenarios.

It is therefore natural to view \eqref{eq:WGCcondition} as a condition for matter on an extremal black hole background to be ``healthy''. We motivate this point of view by applying our condition to extremal rotating BTZ black holes and showing that \eqref{eq:WGCcondition} is satisfied as a consequence of the Null Energy Condition (NEC). This follows from perturbing a BTZ black hole with NEC-satisfying matter holographically dual to a relevant deformation in the CFT. This relevant operator triggers a Renormalization Group (RG) flow along which the central charge monotonically decreases as a consequence of the $c$-theorem. On the black hole side, a decrease in the central charge increases the extremal angular momentum-to-mass ratio, so that a ``spinning'' version of the WGC is satisfied. Although the WGC is normally phrased in terms of black holes charged under $U(1)$ gauge fields,\footnote{One conceptual difference between black holes carrying electric charge versus angular momentum is that spinning black holes are naturally unstable via the Penrose process. On the other hand, extremal electrically charged black holes can provide a large family of stable non-supersymmetric states unless the spectrum is modified from that of pure Einstein-Maxwell theory.} we believe this spinning version of the WGC to also be of interest. 

First, our results suggest a generalization of the Repulsive Force Conjecture (RFC) proposed in \cite{Heidenreich:2019zkl} (based on earlier work in \cite{Palti:2017elp}). If we consider the gravitational and centrifgual forces of a single rotating black hole, we note that a co-rotating object of negligible mass hovering at the event horizon of an extremal rotating black hole experiences an attractive gravitational force that precisely cancels the repulsive centrifugal force. When the spinning WGC is satisfied (but not saturated), the angular momentum-to-mass ratio increases causing the gravitational force to become weaker than the centrifugal force. A similar condition for higher-spin states was studied in \cite{Kaplan:2020ldi}. Second, corrections to the BTZ extremality bound also play an important role in determining the consistency of pure three-dimensional gravity. In \cite{Benjamin:2019stq}, it was observed that the partition function constructed in \cite{Maloney:2007ud,Keller:2014xba} contains a negative density of states in the regime where BTZ black holes are near extremality. One way to cure this pathology is to modify the theory by including additional matter (which could be very heavy) that modifies the BTZ extremality bound and corrects the density of states in a way that guarantees positivity. An alternative resolution has been proposed in \cite{Maxfield:2020ale}. Third, in string theory BTZ black holes can appear as the near-horizon limit of a black string. Upon compactifiying the black string, a charged black hole appears.\footnote{Alternatively, one can use U-duality to map the charged black hole to a BTZ black hole \cite{Skenderis:1999bs}.} If we consider higher-derivative corrections to the black string, one can get constraints on the black hole solution by imposing the spinning WGC in the near-horizon geometry of the black string.

We test this idea by studying a five-dimensional boosted black string. We show that the extremal entropy of the BTZ still matches the four-dimensional entropy after including higher-derivative corrections, but the correction to their respective extremality bounds do not coincide.\footnote{Note that this is not in conflict with the relation between microcanonical entropy (which is not evaluated at zero temperature) and extremality \cite{Hamada:2018dde,Goon:2019faz}.} Instead, they contain complementary information and by imposing both the spinning and charged WGC we obtain positivity bounds on the five-dimensional Wilson coefficients that are stronger than those obtained from the charged WGC alone. The fact that the three-dimensional spinning WGC does not imply the four-dimensional charged WGC, but offers complementary information, agrees with the phenomenon that in theories of gravity with spacetime dimension $d\geq4$, IR consistency cannot completely constrain the sign of corrections to the extremality bound \cite{Hamada:2018dde}. Our finding that positivity bounds can be strengthened by dimensional reduction is also supported by \cite{Cremonini:2020smy}, who, independently of us, study higher-derivative corrections to the extremality bound of five-dimensional black objects and their four-dimensional Kaluza-Klein reductions.

The rest of this paper is organized as follows. In Sec.\ \ref{sec:energyCond} we first loosely motivate why a condition of the form \eqref{eq:WGCcondition} should be true. We then formalize this idea by explicitly deriving the condition using the Iyer-Wald formalism and, as an example, apply our relation to Reissner-N\"ordstrom and BTZ black holes. In Sec.\ \ref{sec:holoRG} we focus on BTZ black holes and show that our derived relation, and therefore a spinning form of the WGC, is satsified as a direct consequence of the holographic $c$-theorem. Finally, in Sec.\ \ref{sec:higherdim} we study higher-derivative corrections to a five-dimensional black string and compute corrections to the extremality bounds of the near-horizon $\text{BTZ}\times S^2$ geometry and the four-dimensional charged black hole that arises after a Kaluza-Klein compactification. Some technical details regarding the Iyer-Wald formalism are reviewed in App.~\ref{app:covphasespace} and the explicit form of the higher-deriative corrected black string solution is described in App.~\ref{app:explicit5Dstring}.

\section{Condition on the stress tensor}\label{sec:energyCond}

\subsection{Loose motivation}
Given an extremal charged black hole perturbed by quantum or higher-derivative corrections, it is of interest to understand under what conditions the charge-to-mass ratio increases in a canonical ensemble (fixed temperature and charge), such that the mild form of the WGC is satisified. In this section, we will rewrite the shift to the extremality bound as a condition on the stress tensor that captures these corrections. In \cite{Hamada:2018dde,Loges:2019jzs}, it was explained in great detail (see also \cite{Goon:2019faz}) that corrections to the extremality bound in a canonical ensemble and corrections to the entropy in a microcanonical ensemble (fixed mass and charge) of an extremal black hole are directly related. At least for stationary black holes both corrections are determined by a modification of the same metric function $f(r)$, whose roots give the location of the horizon of a black hole. To have a concrete example in mind, we can think of $f(r)$ as the $rr$ component of the inverse metric in Schwarszchild gauge of a Reissner-Nordstr\"om black hole perturbed by higher-derivative corrections. Schematically, such a black hole is described by the following action.
\be
I = \frac1{16\pi G_d} \int \of ^dx\sqrt{-g}\left(R - \frac14F_{ab}F^{ab} + \alpha_iI_{\rm hd}^{(i)} \right) ~.
\ee
Here $\alpha_i$ are Wilson coefficients and $I_{\rm hd}^{(i)}$ are higher-derivative terms. Fixing mass and charge, we can write the corrections to $f(r)$ as $f(r) = f_0(r) + \delta f(r) $ and the corrected horizon as $r_+ = r_0 + \delta r$. The location of the corrected horizon is now found by solving
\be
f(r_0 + \delta r) \simeq f_0(r_0) + \delta f(r_0) + f_0'(r_0)\delta r+\frac12 f_0''(r_0)\delta r^2 + (\ldots)  = 0 ~,
\ee
where we treated the correction as a small perturbation. For an extremal black hole $f_0(r_0) = f'_0(r_0)=0$ and the shift in the horizon is given by
\be
\delta r = \pm \sqrt{\frac{-2\delta f(r_0)}{f''(r_0)}} ~,
\ee
when $\delta f(r_0)$ and $f''(r_0)$ are both non-vanishing. Because $f''(r_0) > 0$, the singularity of the solution is only cloaked behind a horizon when $\delta f(r_0) \leq 0$, which shifts the outer horizon positively (or leaves it uncorrected). The Wald entropy of the corrected black hole is now given by
\be
S = \frac{A(r_0+\delta r)}{4G_d} + \delta S_\text{w} ~,
\ee
where the first term is the Bekenstein-Hawking entropy and the second term contains a modification to the Bekenstein-Hawking formula due to higher-derivative corrections. Because the horizon shift of an extremal black hole scales as  $\delta r \sim {\cal O}(\sqrt{\alpha_i})$, we find that
\be
S = \frac{A(r_0+\delta r)}{4G_d} + {\cal O}(\alpha_i) ~.
\ee
The leading piece of the black hole entropy is simply given by the Bekenstein-Hawking formula evaluated on the corrected horizon. Equivalently, we can also consider the corrections in a canonical ensemble. In that case, a microcanonical increase in the horizon manifests itself as a decrease to the ADM mass, increasing the charge-to-mass ratio \cite{Hamada:2018dde,Loges:2019jzs}. Thus, when working in a microcanonical ensemble the WGC can be understood as the statement that singularities present in the uncorrected spectrum should be cloaked behind a horizon after including corrections. However, we should stress that this does not imply that the mild form of the WGC follows from the Weak Cosmic Censorship Conjecture. Here we are comparing two different black holes (one with and one without higher-derivative corrections) to each other and not having a positive real shift of the outer horizon, i.e. $\delta r<0$, at odds with the WGC only implies that an extremal black hole in the uncorrected theory is not a regular solution in the corrected theory.

\begin{figure}
\centering
\begin{tikzpicture}
\filldraw[fill=gray!20, line width=2 pt] (-6,0) circle [radius=1.5] node {$Q,M$};
\filldraw[fill=gray!20, line width=2 pt] (0,0) circle [radius=1.5] node {$Q,M'$};
\draw[dotted,line width=1 pt] (0,0) circle [radius=1.7];
\draw[line width=2pt,color=gray] (0,0) circle [radius=2];
\node at (0,2.4) {$m=M-M'$};
\end{tikzpicture}
\caption{Cartoon representation of WGC-satisfying corrections to a black hole. On the left we have a black hole with charge $Q,$ and mass $M$, where $M$ is the ADM mass, in the unperturbed theory. On the right, we add matter to the theory, imagining that it contributes as an uncharged shell (the outer gray ring) with mass $m$. Keeping the ADM mass fixed, we have $m=M-M'$, where $M'$ is the new mass of interior region. The WGC dictates that the singularity remains cloaked in the perturbed solution, so that $M'\geq M$ and $m\leq 0$. The dotted line represents the horizon of the perturbed geometry with $m< 0$.}\label{fig:shell}
\end{figure}
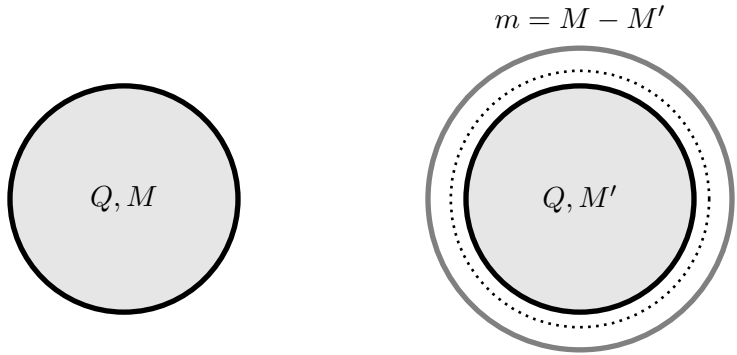

When this correction to the extremality bound is induced by additional matter (for example heavy matter that is integrated out, generating higher-derivative terms), it is natural to expect that whenever that matter is ``healthy'' it leads to a correction compatible with the WGC. Indeed, in the set-up considered in \cite{Hamada:2018dde,Loges:2019jzs} this is precisely what happens; unitarity and causality imply WGC-compatible signs of the Wilson coefficients. We will now phrase this healthiness in terms of the condition on the stress tensor to which we alluded earlier. To first gain some intuition, we imagine that the effect of the additional matter is to introduce a shell outside the horizon of an extremal charged black hole. For simplicity, we suppose the shell is uncharged and has a mass $m$: see Fig.\ \ref{fig:shell}.
Before introducing the shell, the extremal black hole has a charge-to-mass ratio of $Q/M=1$ (in appropriate units), where $M$ is the ADM mass. Keeping the ADM mass and charge fixed, we now introduce the shell. The charge-to-mass ratio is now given by
\begin{align}
    \frac{Q}{M}=\frac{Q}{M'+m}=1 ~.
\end{align}
As explained, the WGC now dictates that the resulting state does not contain a naked singularity, which means that the $Q/M' \leq 1$. Since the ADM mass and charge are held fixed this requires $m\leq 0$. In terms of the matter stress tensor this condition reads
\be \label{eq:almostcorrect}
m = \int_\Sigma \mathrm{d}^{d-1}x \,\sqrt{h}\,T_{ab}\xi^an^b \leq 0 ~.
\ee
Here $\Sigma$ is a Cauchy slice of constant Killing time $t$ with an induced metric $h_{ab}$ and a unit normal vector $n^a$. We are interested in stationary black holes, so $\xi^a$ is a timelike Killing vector. So we see that, at least in this example, we can rephrase the correction to the extremality bound as a covariant condition on the stress tensor.

While this simple setup gives some useful intuition, it also has its shortcomings. In particular, if the two-derivative action includes a graviton and gauge field as massless degrees of freedom, additional matter fields also backreact on the gauge field and must be taken into account. We now show that the correct condition \eqref{eq:WGCcondition} also takes into account a correction to the stress tensor of the gauge field.

\subsection{Deriving the general relation}\label{sec:shiftGeneral}
Now that we have motivated the WGC as a condition on a stress tensor, we make this intuition precise by rewriting the corrections to the horizon of an extremal black hole as an integral of the stress tensor. To do so, it will be useful to employ the covariant phase space formalism of Iyer and Wald, which we review in App. \ref{app:covphasespace}. Viewing the Lagrangian as a $d$-form, we consider Einstein-Maxwell theory, possibly with a cosmological constant:
\be
{\bf L} = \frac1{2\kappa^2}\left(R-2\Lambda\right)\epsilon -\frac12F\wedge \star F~.
\ee
Here $\kappa^2 = 8\pi G_d$ and $\epsilon$ is the volume form on the $d$-dimensional background. As explained in the appendix, for any infinitesimal diffeomorphism parametrized by $\xi$ or gauge transformation parametrized by $\lambda$ we can construct a Hamiltonian that obeys a conservation equation that is satisfied on-shell. We now consider an off-shell variation of the Hamiltonian. We then find (see \eqref{eq:appconservation}) the following conservation equation:
\be \label{eq:variationconservation}
\of \delta{\bf H} = -2\delta\star(E_\text{g}\cdot\xi) -(\iota_\xi A+\lambda)\of\star \delta F  ~.
\ee
The left-hand side is a variation of the exterior derivative of the Hamiltonian and the right-hand side contains a term $E_\text{g}$ that captures the gravitational equations of motion and a second term that arises from a variation with respect to the gauge field $A$. Because the background satisfied Einstein's equations, we can rewrite the first term as
\be
-2\delta\star(E_\text{g}\cdot\xi) = -\star \delta T_{ab}\xi^b\of x^a ~.
\ee
Now let us consider a black hole (not necessarily in asymptotically flat space). Integrating \eqref{eq:variationconservation} over a Cauchy slice $\Sigma$ of constant Killing time $t$ located somewhere between the (outer) horizon and spatial infinity, we can use Stokes' theorem to write
\be 
\delta H = \int_{S^{d-2}_\infty}\delta{\bf H} = \int_{S^{d-2}_\text{hor}}\delta{\bf H} + \int_\Sigma \mathrm{d}^{d-1}x\,\sqrt{h}\,\left(\delta T_{ab} +F_{ac}\delta F_b^{\,\,\,c}\right)n^a\xi^b ~.
\ee
To arrive at this form, we picked a gauge in which $\iota_\xi A+\lambda$ vanishes at the horizon and we assumed that $\delta F_{ab}$ dies off sufficiently fast at infinity. Here $h_{ab}$ is the induced metric on $\Sigma$ and $n^a$ its unit normal vector. The second integral on the right-hand side consists of two terms, which arise from varying both the metric and gauge field. The sum of both contributions can be thought of as an effective stress tensor
\be
\delta T_{ab}^{\rm eff} = \delta T_{ab} + F_{ac}\delta F_b^{\,\,\,c} ~,
\ee
and we arrive at the following relation
\be \label{eq:conservation}
\left(\int_{S^{d-2}_\infty} - \int_{S^{d-2}_\text{hor}}\right)\delta{\bf H} = \int_\Sigma \mathrm{d}^{d-1}x\,\sqrt{h}\,\delta T_{ab}^{\rm eff}n^a\xi^b ~.
\ee
As we will see next, when we specify a black hole background the first integral on the left-hand side becomes proportional to the asymptotic charges of the black hole and the second integral gives the correction to the horizon. At fixed asymptotic charges, we then find a identity relating the shift of the black hole horizon to the stress tensor, reformulating the WGC as a condition on the stress tensor. A non-covariant version of this relation already appeared in \cite{Kats:2006xp} in the context of the four-dimensional Reissner-N\"ordstrom black hole. Here we considered Einstein-Maxwell theory, but a generalization to a more general theory with stationary black hole solutions is straightforward.

\subsubsection{BTZ black hole}
The first example we look at is pure three-dimensional gravity described by the following Lagrangian.
\be
{\bf L} = \frac1{2\kappa^2}\left(R-2\Lambda\right)\epsilon ~.
\ee
On a constant negative-curvature background, $\Lambda=-1/\ell^2$, a particular solution of Einstein's equations is given by the BTZ black hole which has the metric
\be \label{eq:BTZmetric}
\of s^2 = -N(r)^2\,\of t^2 + N(r)^{-2}\,\of r^2 + r^2\big(\of \phi+N^\phi(r)\,\of t\big)^2 ~,
\ee
with
\be
N(r)^2 = \frac{(r^2-r_+^2)(r^2-r_-^2)}{\ell^2r^2} ~, \quad N^\phi(r) = \frac{r_+r_-}{\ell r^2} ~.
\ee
The inner and outer horizon are given by $r_\pm$ and the mass $M_3$ and angular momentum $J_3$ can be written as
\be \label{eq:BTZparameters}
M_3 = \frac{r_+^2+r_-^2}{8G_3\ell^2} ~, \quad J_3 = \frac{r_+r_-}{4G_3\ell} ~.
\ee
To construct the Hamiltonian associated with an infinitesimal diffeomorphism, we first have to compute the symplectic potentials and Noether charges (see \eqref{eq:symplecticg} and \eqref{eq:charges}). For the Killing vector $\xi=\partial_t$ we then find the following expressions.
\begin{align}
{\bf Q}_{\partial_t} &= \frac1{2\kappa^2}\left(-r^3N^\phi(r) N'^{\phi}(r) +2r N(r)N'(r)\right)\of\phi + \left(\dots\right)\of t  ~, \\
\iota_{\partial_t}{\bf \Theta} &=\frac1{2\kappa^2}\left(-r^3\delta N^\phi(r)N'^\phi(r)+2r\delta N(r)N'(r) + 2N(r)\big(\delta N(r)+r\delta N'(r)\big)\right)\of \phi \nn \\
&\qquad+ \left(\ldots\right)\of t ~.\nn
\end{align}
We only displayed the terms proportional to $\of \phi$ because the other terms will drop out of the integral of interest. Taking variations with respect to the metric functions and using \eqref{eq:varHamil} we find the variation of the Hamiltonian for the timelike Killing vector.
\be
\delta {\bf H}_{\partial_t} =-\frac{\of\phi}{2\kappa^2}\left(2N(r)\delta N(r)+r^3N^\phi(r)\delta N'^\phi(r)\right) ~.
\ee
The conserved charge associated with with this Hamiltonian is of course the mass:
\be
\delta H_{\partial_t} = \int_{S^1_\infty}\delta {\bf H}_{\partial_t} =  \delta M_3 ~.
\ee
Similarly, the Noether charge and symplectic potential for the Killing vector $\partial_\phi$ are
\begin{align}
{\bf Q}_{\partial_\phi} &= -\frac1{2\kappa^2}r^3N'^\phi(r) \of\phi + \left(\ldots\right)\of t  ~, \\
\iota_{\partial_\phi}{\bf \Theta} &=\left(\ldots\right)\of t ~,\nn
\end{align}
such that
\be
\delta {\bf H}_{\partial_\phi} =   -\frac1{2\kappa^2}r^3\delta N'^\phi(r) \,\of\phi ~.
\ee
The associated conserved charge is the angular momentum:
\be
\delta H_{\partial_\phi} = \int_{S^1_{\infty}}\delta {\bf H}_{\partial_\phi} = \delta J_3 ~.
\ee
In \cite{Maxfield:2019hdt} it was observed that the integral of the Hamiltonian variation associated to the Killing vector $K = \partial_t - \Omega\partial_\phi$ (where $\Omega = \frac{r_-}{\ell r_+}$ is the angular potential) over the horizon is directly proportional to the variation of $N(r_+)^2$, whose roots determine the location of the horizon:
\be
\int_{S^1_\text{hor}} \delta {\bf H}_{K} = -\frac1{8 G_3} \delta (N(r_+)^2) ~.
\ee
Notice that for this Killing vector, the horizon is a Killing horizon. We can now relate the shift in the horizon to the stress tensor and the conserved charges by making use of \eqref{eq:conservation}, leading to
\be
 {-\frac1{8 G_3}} \delta (N(r_+)^2) =  \delta M_3 - \Omega\delta J_3  -\int_\Sigma \mathrm{d}^{2}x\,\sqrt{h} \,\delta T_{ab}n^aK^b ~.
\ee
Hence, the horizon shift for an extremal black hole at fixed charges ($\delta M_3=\delta J_3=0$) is determined by
\be \label{eq:BTZshift}
\frac1{8 G_3} \delta (N(r_+)^2) =  \int_\Sigma \mathrm{d}^{2}x\,\sqrt{h}\,\delta T_{ab}n^aK^b  ~.
\ee
As explained, a positive (or absent) shift of the horizon requires $\delta(N(r_+)^2) \leq 0$ which leads to
\be \label{eq:BTZintegratedcondition}
\int_\Sigma \mathrm{d}^{2}x\,\sqrt{h}\,\delta T_{ab}n^aK^b \leq 0 ~.
\ee
Given this condition it is now straightforward to determine whether a particular correction to the BTZ background increases the horizon in a microcanonical ensemble (fixed $M_3$ and $J_3$), which determines the extremality bound in a canonical ensemble (fixed temperature and $J_3$). In \cite{Maxfield:2019hdt} for example, this relation has been employed to compute the correction to the extremality bound induced by the one-loop stress tensor of a massive scalar field.

Here, we are interested in computing higher-derivative corrections that are generated upon integrating out heavy matter and we will perturb the BTZ black hole by the leading gravitational corrections in a derivative expansion,
\be
\Delta{\bf L} = m^{-1}\left(\alpha_1 R^2 + \alpha_2R_{ab}R^{ab}\right)\epsilon ~.
\ee
The mass scale $m$ in front, which obeys $\ell m\gg 1$, is chosen such as to make the coefficients $\alpha_1,\alpha_2$ dimensionless. Varying the higher-derivative Lagrangian with respect to the metric, we find that the stress tensor is given by
\begin{align}
\delta T_{ab} &= \alpha_1m^{-1}\left(-4R_{ab}R+g_{ab}R^2+4\nabla_a\nabla_b R-4g_{ab}\square R\right)\\
&\quad +\alpha_2m^{-1}\left(g_{ab}R_{cd}R^{cd} +2\nabla_a\nabla_b R-4R_{cadb}R^{cd} -2\square R_{ab}-\square R g_{ab} \right) ~. \nn
\end{align}
Evaluated on the BTZ background the stress tensor is
\be
\delta T_{ab} = -\frac{4(3\alpha_1+\alpha_2)}{\ell^4m}g_{ab} ~.
\ee
Plugging this into \eqref{eq:BTZshift}, we find a divergent integral. This divergence can be blamed on the fact that higher-derivative corrections do not fall off going to the boundary in three-dimensional gravity. In the context of holography, it is well known that the proper way to regulate the stress tensor in three-dimensional asymptotically anti-de Sitter space is to subtract the contribution of the cosmological constant \cite{Balasubramanian:1999re}. In our case, this contribution is
\be
\delta T_{ab}^{(0)} = -\frac{4(3\alpha_1+\alpha_2)}{\ell^4m}g_{ab}^{(0)} ~,
\ee
where $g_{ab}^{(0)}$ is the metric of empty AdS. We can now perform the integral to find the finite result
\be
\delta (N(r_+)^2) =-\frac{32G_3\pi r_+^2}{\ell^4m}(3\alpha_1+\alpha_2) ~.
\ee
It is straightforward to check by explicitly solving the corrected Einstein equations that this result, obtained by the regularization procedure described above, indeed yields the correct modification to the geometry. The horizon is shifted positively (or receives no corrections) when $3\alpha_1+\alpha_2 \geq 0$. 

\subsubsection{Reissner-Nordstr\"om black hole}
We now repeat the above calculation for electrically charged Reissner-Nordstr\"om black holes in asymptotically flat four-dimensional space. These black holes are solutions to the following Lagrangian:
\be
{\bf L} = \left(\frac1{2\kappa^2}R\right)\epsilon -\frac12F\wedge \star F~.
\ee
The line element of the Reissner-Nordstr\"om black hole solution is given by
\begin{align}
    \of s^2 &= -f(r)\,\of t^2 + f(r)^{-1}\,\of r^2 + \rho(r)^2(\of \theta^2+\sin^2\theta \,\of \phi^2) ~,\\
    A &= -\Phi(r)\,\of t ~,\nn
\end{align}
with
\begin{equation}
    f(r) = \frac{(r-r_+)(r-r_-)}{r^2} ~, \quad \rho(r) = r ~, \quad \Phi(r) = \frac{Q}{4\pi r} - \Phi_+ ~,
\end{equation}
where it is useful to choose a gauge in which $A_t$ vanishes at the horizon: $\Phi_+$ gives the difference in electric potential between the horizon and infinity.\footnote{Subtleties regarding gauge invariance of a choice of potential have been pointed out in \cite{Elgood:2020svt,Elgood:2020mdx,Elgood:2020nls}. Because \eqref{eq:electricinequality} is independent of $\Phi_+$ our final result is gauge invariant.} The mass and electric charge are given by
\be
M_4 = \frac{r_++r_-}{2G_4} ~, \quad Q^2 = \frac{4\pi r_+r_-}{G_4}  ~.
\ee
We now need to consider the Hamiltonian that generates the flow of the timelike Killing vector $\partial_t$, for which the black hole horizon is a Killing horizon, and the Hamiltonian for the gauge transformation $A \to A + \of\lambda$. Proceeding as before, we obtain explicit expressions for the Noether charge and symplectic potential (using \eqref{eq:symplecticA} and \eqref{eq:charges}).
\begin{align}
    {\bf Q}_{\partial_t} &=\frac1{2\kappa^2} f'(r)\rho(r)^2\sin{\theta}\, \of\theta\wedge\of\phi  ~, \\
    \iota_{\partial_t}{\bf \Theta}_\text{g} &= \frac1{2\kappa^2}\left(\frac{2\rho'(r)}{\rho(r)}\delta f(r) + \delta f'(r) + \frac{4f(r)}{\rho(r)}\delta\rho'(r)\right)\rho(r)^2\sin\theta\, \of\theta\wedge\of\phi~.\nn
\end{align}
Similarly, the charge associated with gauge transformations is
\be
{\bf Q}_\lambda = -\Phi(r)\Phi'(r)\rho(r)^2\sin\theta\, \of\theta\wedge\of\phi ~.
\ee
The variation of the Hamiltonians are (using \eqref{eq:varHamil} and \eqref{eq:varHamil2})
\begin{align}
    \delta{\bf H}_{\partial_t} &= -\frac1{8\pi G_4}\left[\frac{\rho'(r)}{\rho(r)} \delta f(r) + \frac{2f(r)}{\rho(r)}\delta\rho'(r) - \frac{f'(r)}{\rho(r)}\delta\rho(r)\right]\rho(r)^2\sin\theta\,\of\theta\wedge\of\phi ~, \\
    \delta{\bf H}_{\lambda} &= -\left[\Phi(r)\delta \Phi'(r) + \frac{2\Phi(r)\Phi'(r)}{\rho(r)}\delta\rho(r)\right]\rho(r)^2\sin\theta\, \of\theta\wedge\of\phi\nn ~,
\end{align}
and the corresponding conserved charges are
\begin{align}
\delta H_{\partial_t} &= \int_{S^2_\infty}\delta{\bf H}_{\partial_t } = \delta M_4  ~, \\
\delta H_{\lambda} &= \int_{S^2_\infty}\delta{\bf H}_{\lambda } = {-\Phi_+}\delta Q \nn ~.
\end{align}
We now use \eqref{eq:conservation} to find the shift in the horizon due to additional contributions to action (such as higher-derivative terms):
\be
-\frac{r_+\delta f(r_+)}{2G_4} = \delta M  - \Phi_+\delta Q - \int_\Sigma \mathrm{d}^3x \,\sqrt{h}\,\left(\delta T_{ab} +F_{ac}\delta F_b^{\,\,\,c}\right)n^a\xi^b ~.
\ee
Thus, at fixed charges ($\delta M_4=\delta Q=0$) we obtain
\be \label{eq:RNshift}
\frac{r_+\delta f(r_+)}{2G_4} = \int_\Sigma \mathrm{d}^3x\,\sqrt{h}\,\delta T^{\rm eff}_{ab}n^a\xi^b ~,
\ee
where the effective stress tensor is defined as
\be
\delta T^{\rm eff}_{ab} = \delta T_{ab} +F_{ac}\delta F_b^{\,\,\,c} ~.
\ee
Because a positive shift (or no corrections) of the horizon requires $\delta f(r_+)\leq 0$ we find that the mild form of the WGC can be rewritten as a condition on the effective stress tensor.
\be \label{eq:electricinequality}
\int_\Sigma \mathrm{d}^3x\,\sqrt{h}\,\delta T^{\rm eff}_{ab}n^a\xi^b \leq 0 ~.
\ee
We will now show how this relation can be employed by perturbing the Reissner-Nordstr\"om black hole with the following higher-derivative corrections:
\be
\Delta{\bf L} = \left(\frac{a_1}{4}(F_{ab}F^{ab})^2 +\frac{a_2}{2}F_{ab}F_{cd}W^{abcd} \right)\epsilon ~.
\ee
We are interested in purely electric solutions, so we omit a term of the form $F_{ab}F^{bc}F_{cd}F^{da}$ which can be written as a multiple of $(F_{ab}F^{ab})^2$. Dimensionless coefficients are defined by $b_1 = a_1/\kappa^4$ and $b_2 = a_2/\kappa^2$. For an electrically charged Reissner-Nordstr\"om these are the most general higher-derivative corrections up to four derivatives.

As we saw, the effective stress tensor $\delta T_{ab}^{\rm eff}$ contains explicit terms $\delta T_{ab}$ that arise from varying $\Delta {\bf L}$ with respect to the metric as well as implicit corrections $F_{ac}\delta F^c_{\,\,\,b}$ that capture a modification of Maxwell's equations. The corrected Maxwell equations are given by
\begin{align} \label{eq:MaxwellMod}
\nabla_bF^{ab} &= 2\nabla_b\left(2b_1\kappa^4F_{cd}F^{cd}F^{ab} + b_2\kappa^2W^{abcd}F_{cd} \right) ~.
\end{align}
This is solved by
\be
F = \left(- \frac{Q}{4\pi r^2} + \frac{4 b_1 G_4^2 Q^3}{\pi  r^6} + \frac{4 b_2 G_4^2 \left(Q^3-4 \pi  M_4 Q r\right)}{\pi  r^6}\right)\of t\wedge\of r ~.
\ee
The explicit corrections are given by \cite{Kats:2006xp}
\begin{equation} \label{eq:4DStressMod}
\begin{aligned}
\delta T_{ab} &= \frac{b_1}{4}\kappa^4\left(g_{ab}(F_{cd}F^{cd})^2-8F_{cd}F^{cd}F_a^{\,\,\,e}F_{be}\right) \\
&\quad {}+ \frac{b_2}{2}\kappa^2\bigg[g_{ab}R_{cdef}F^{cd}F^{ef}-6F_{cb}F^{de}R^c_{\,\,\,ade}-4\nabla_d\nabla_c\big(F^c_{\,\,\,a}F^d_{\,\,\,b}\big) \\
&\qquad {}-2g_{ab}R^{cd}F_{ce}F_{d}^{\,\,\,e} +8R_{bc}F_{ad}F^{cd} + 4 R^{cd}F_{ca}F_{db} +2 g_{ab} \nabla_c\nabla_d\big(F^c_{\,\,\,e}F^{de}\big) \\
&\qquad {}-4\nabla_c\nabla_b\big(F_{ad}F^{cd}\big)+2\square\big(F_{ac}F_{b}^{\,\,\,c}\big)+\frac13g_{ab}RF_{cd}F^{cd} \\
&\qquad {}-\frac43RF_{a}^{\,\,\,c}F_{bc}-\frac23F_{cd}F^{cd}R_{ab}+\frac23\nabla_a\nabla_b\big(F_{cd}F^{cd}\big) -\frac23g_{ab}\square\big(F_{cd}F^{cd}\big) \bigg] \,.
\end{aligned}
\end{equation}
Adding both corrections and performing the integral over $\Sigma$ and using \eqref{eq:RNshift} we find in the extremal limit
\be
\delta f(r_+) = - \frac{64\pi^2(2b_1-b_2)}{5Q^2} ~.
\ee
A positive (or absent) shift of the horizon requires $2b_1-b_2 \geq 0$, which matches \cite{Kats:2006xp,Hamada:2018dde}.

\section{Spinning WGC from Holographic RG}\label{sec:holoRG}
In the previous section, we rephrased WGC-satisfying corrections to the extremality bound as an integrated condition on an effective stress tensor and gave two examples where the corrections to the stress tensor arose from higher-derivative terms, but we could also have considered quantum corrections. In \cite{Maxfield:2019hdt}, this relation was used to relate the one-loop stress tensor of a massive scalar field to a correction of the BTZ extremality bound. To compute the effect of quantum corrections, one simply replaces the classical value of the stress tensor by its expectation value. This correctly gives the shift in the horizon radius as long as the semi-classical approximation is valid. 

In light of a spinning WGC we would like to understand whether there is a general principle behind positivity of the horizon shift. In this section, we show that this is the case for a particular class of corrections to the BTZ black hole. In particular, when a BTZ black hole is perturbed by a relevant deformation, this triggers a holographic RG flow. When the NEC is satisfied along the flow, the central charge of the dual CFT decreases by virtue of the $c$-theorem \cite{Zamolodchikov:1986gt}. This implies that when we reach a fixed point in the IR, that theory includes BTZ black holes with angular momentum-to-mass ratios exceeding those of unperturbed black holes, so that a spinning version of the WGC is satisfied.

The holographic RG \cite{deBoer:1999tgo}, provides a systemic way of computing CFT correlation functions from the on-shell gravitational action at fixed radial coordinate in the context of AdS/CFT. The radial coordinate in the bulk is identified as an energy scale in the CFT and moving from the boundary of AdS into the bulk describes an RG flow from the UV to the IR in the boundary theory. For our purposes, we will consider the following three-dimensional action on an AdS background perturbed by purely gravitational four-derivative operators:
\be \label{eq:3Dhdcorrected}
I = \int \mathrm{d}^3x\,\sqrt{-g}\,\left(\frac1{2\kappa^2}\left(R+\frac2{\ell^2}\right) + \alpha_1\ell R^2 + \alpha_2\ell R_{ab}R^{ab}\right) ~.
\ee 
For now we focus on these particular higher-derivative corrections, but our method can be easily generalized to additional terms as well. In the context of the holographic RG we can think of this action as an effective field theory in the IR whose higher-derivative corrections parametrize the effect of modes that have been integrated out along the flow. In three dimensions, the Ricci tensor is proportional to the metric which implies that the higher-derivative corrected action still has a BTZ solution described by the metric \eqref{eq:BTZmetric}. The effect of adding higher-derivative corrections is to shift the central charge of the dual CFT$_2$ from its Brown-Henneaux \cite{Brown:1986nw} value. The corrected central charge can easily be determined using $c$-extremization \cite{Kraus:2005vz}. One defines a $c$-function given by
\be
c(\ell) = \frac{3\ell^2}{8G_3}{\cal L}_3 ~,
\ee
and extremizes this with respect to $\ell$ to find the central charge. Here ${\cal L}_3$ is the Euclidean Lagrangian. In our case, we have
\be
{\cal L}_3 = -R-\frac{2}{\ell^2} - 2\kappa^2\ell\left(\alpha_1 R^2+\alpha_2 R_{ab}R^{ab}\right) ~.
\ee
Extremizing the $c$-function, we obtain
\be \label{eq:centralcharge}
c = \frac{3\ell}{2G_3}\left(1-\frac{6\kappa^2(3\alpha_1+\alpha_2)}{\ell}\right) ~.
\ee 
From Einstein's equations one finds that the higher-derivative terms are proportional to the cosmological constant, so we can also absorb them into the AdS length. Explicitly,
\be
\alpha_1 R^2+\alpha_2R_{ab}R^{ab} = \frac{12}{\ell^4}\left(3\alpha_1+\alpha_2\right) ~.
\ee
Thus, we can write \eqref{eq:3Dhdcorrected} equivalently as
\be
I = \frac1{16\pi G_3}\int \of ^3x\,\sqrt{-g}\,\left(R+\frac{2}{L^2}\right) ~,
\ee
with
\be \label{eq:AdSlengthcorr}
L = \ell - 6\kappa^2(3\alpha_1+\alpha_2) ~.
\ee
Since we removed the higher-derivative corrections, this action has BTZ solutions with an AdS length $L$ and the central charge of the dual CFT is now just given by the Brown-Henneaux value $c=3L/2G_3$. Indeed, using \eqref{eq:AdSlengthcorr} the central charge is still given by
\be\label{eq:Ll}
c=\frac{3L}{2G_3} =\frac{3\ell}{2G_3}\left(1-\frac{6\kappa^2(3\alpha_1+\alpha_2)}{\ell}\right) ~.
\ee
We therefore see that on-shell, higher-derivative corrections in three dimensions can equivalently be understood as an uncorrected theory with a modified AdS length \cite{Witten:2007kt}.

The correction to the central charge modifies the entropies and extremality bound of black holes in the theory. The mass and angular momentum given in \eqref{eq:BTZparameters} are related to the excitation levels of the dual CFT by the standard relations \cite{Kraus:2006wn}
\be\label{eq:MJh}
M_3\ell = h+\bar h - \frac{c}{12} ~, \quad J_3 = h-\bar h ~.
\ee
In terms of the excitation levels we can write the extremality bound as
\be\label{eq:newExt}
\bar h \geq  \frac{c}{24} ~.
\ee
To derive the change in extremality bound in a canonical ensemble and the change in entropy in a microcanonical ensemble we find it useful to use a thermodynamic approach. The Euclidean action is given by
\be
I_E = -\int \of^3x\sqrt{g}\left(\frac1{2\kappa^2}\left(R+\frac2{\ell^2}\right) + \alpha_1\ell R^2 + \alpha^2\ell R_{ab}R^{ab}\right) - \frac1{\kappa^2}\oint \of^2x\sqrt{h}\left(K-K_0\right)
\ee
where we supplemented the bulk action by a Gibbons-Hawking-York boundary term, defined at the boundary at $r\to\infty$, and a counterterm $K_0$ to make the on-shell action finite. The counterterm that removes the divergence of $I_E$ when $r\to\infty$ is given by
\be
K_0 = \frac1\ell + \frac{6\kappa^2(3\alpha_1+\alpha_2)}{\ell^2} ~.
\ee
Using this, the on-shell action is given by
\be
I_E = \frac{\pi  \beta  \left(r_-^2-r_+^2\right)}{\kappa ^2 \ell^2} -\frac{6 \pi  \beta  \left(r_-^2-r_+^2\right) (3 \alpha_1+\alpha_2)}{\ell^3} ~.
\ee
Here $\beta$ is the inverse temperature of the black hole. It is well known that, even in the presence of higher-derivative corrections, the Euclidean action can be written in terms of the Gibbs free energy $G$ as \cite{Reall:2019sah}
\be
I_E = \beta G = \beta\left(M_3-TS-\Omega J_3\right) ~.
\ee
Here $\Omega=\frac{r_-}{\ell r_+}$ denotes the angular potential and $S$ is the entropy. We can now evaluate the Euclidean action in a grand canonical ensemble by writing $I_E = I_E(T,\Omega)$, where $T=\beta^{-1}$. The different thermodynamic quantities are given by
\begin{align} \label{eq:3Dthermo}
S &= -\left(\frac{\partial G}{\partial T}\right)_{\Omega} ~, \quad J_3 = -\left(\frac{\partial G}{\partial \Omega}\right)_{T} ~,
\end{align}
and the mass is
\be \label{eq:3Dmass}
M_3 = G + TS+\Omega J_3 ~.
\ee
We start by computing the extremal entropy by expressing $S(\Omega,T)$ as $S(T,M_3)$ and take $T\to0$. We then find
\be
\left.S\right|_{T=0} = \ell\pi\sqrt{\frac{M_3}{G_3}}\left(1-\frac{48\pi G_3(3\alpha_1+\alpha_2)}{\ell}\right) ~.
\ee
We now evaluate the mass in a canonical ensemble by expressing it in terms of $T$ and $J_3$. In the limit $T\to 0$ we obtain
\be
M_3  = \frac{J_3}{\ell}\left(1 -\frac{48\pi G_3(3\alpha_1+\alpha_2)}{\ell} \right) ~.
\ee
Thus, the extremality bound is modified as
\be
\frac{J_3}{\ell M_3} \leq 1+\frac{48\pi G_3(3\alpha_1+\alpha_2)}{\ell} ~.
\ee
Finally, we are also interested in the microcanonical entropy (fixed $M_3$ and $J_3$). Expanding the canonical expression of the mass for small temperature, we find that the $z=J_3/(\ell M_3)=1$ state has a non-zero temperature of the form
\be
\left.T\right|_{z=1} = \frac{16G_3}{\ell^2}\sqrt{\frac{3J_3(3\alpha_1+\alpha_2)}{\pi}}~.
\ee
At fixed $M_3$ and $J_3$ the correction to the extremal black hole entropy is given by
\be
\left.S\right|_{z=1} = \pi\ell\sqrt{\frac{M_3}{G_3}}\left(1+\sqrt{\frac{48\pi G_3}{\ell}(3\alpha_1+\alpha_2)}\right) ~.
\ee
We see that a positive shift of the angular momentum-to-mass ratio of an extremal BTZ black hole increases the microcanonical entropy and corresponds to a decrease of the central charge. When the correction to the central charge is generated by the higher-derivative corrections in \eqref{eq:3Dhdcorrected} a negative (or absent) shift in the central charge requires $3\alpha_1+\alpha_2 \geq0$.

\begin{figure}[t]
\centering
\includegraphics[scale=1.0]{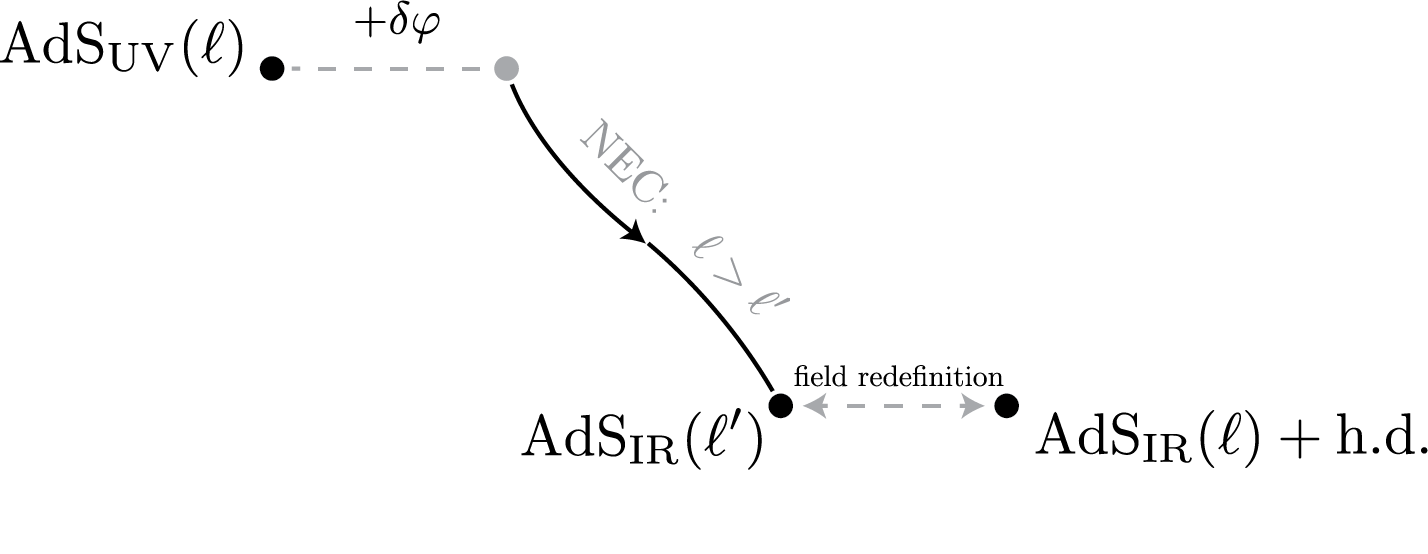}
\caption{If we perturb an AdS space with AdS length $\ell$ by a relevant deformation $\delta\varphi$, this triggers an RG flow until we reach an IR fixed point describing an AdS space with length $\ell'$. When $\delta\varphi$ satisfies the NEC, $\ell > \ell'$. By a field redefinition, the action of the IR AdS space can be related to an AdS space with length $\ell$ and higher-derivative corrections.}
\label{fig:RGflow}
\end{figure}

It is now straightforward to argue that when we perturb a BTZ black hole by a relevant perturbation, the central charge decreases (or is uncorrected) along the flow such that $3\alpha_1+\alpha_2\geq 0$ and a spinning WGC is obeyed. Our starting point is a purely three-dimensional gravity theory with a Brown-Henneaux central charge. Then, we perturb this theory by some matter field that is holographically dual to a relevant operator. This will trigger a holographic RG flow until we reach a fixed point in the IR, which corresponds to a CFT perturbed by an irrelevant deformation. The gravitational dual of this theory has BTZ solutions with an AdS length (and central charge) that is smaller than the one in the unperturbed theory by virtue of the $c$-theorem: see Fig.\ \ref{fig:RGflow}.
So whenever the higher-derivative corrections in \eqref{eq:3Dhdcorrected} arise in the IR as a consequence of a relevant perturbation, $3\alpha_1+\alpha_2 \geq 0$. Next, we will illustrate this behaviour when the relevant perturbation is a scalar field.

Although it is convenient to assume that the UV CFT is dual to pure Einstein gravity, such that its central charge takes the Brown-Henneaux form, this is strictly speaking not necessary. As long as the $c$-theorem is obeyed, it is guaranteed that the IR central charge is smaller than the central charge of the UV fixed point. In this sense, the three-dimensional spinning form of the WGC is insensitive to the UV, as long as there exists a black hole with which to compare the extremality bound.

\subsection{Example: Scalar perturbation}
We now give an explicit example of a holographic RG flow where the relevant perturbation is a scalar field. Because BTZ black holes are related to empty AdS by a modular transformation, we find it convenient to describe a flow between two AdS spaces. A modular transformation does not modify the central charge and the AdS flow is therefore sufficient to show that the central charge decreases. Of course, one could also consider a direct flow between two BTZ black holes as in \cite{Hotta:2008xt}, which is technically more involved. As expected, those results are also in agreement with the $c$-theorem.

To describe the flow, it will be useful to take the following domain-wall ansatz for the metric.
\be
\of s^2 = \of \rho^2 + e^{2A(\rho)}\left(-\of t^2 + \of x^2\right) ~,
\ee
where we take $x\sim x+2\pi$. Empty AdS space (with a compactified $x$-coordinate) corresponds to $A(\rho) = \rho/\ell$. We now perturb the Einstein-Hilbert action by a scalar field $\phi$, writing
\be
I = \int \of ^3x\sqrt{-g}\left(\frac1{2\kappa^2}R -\frac12\partial_a\phi\partial^a\phi - V(\phi) \right) ~.
\ee
Taking the scalar field to depend only on the radial coordinate $\rho$, we can write the action in the following form:
\begin{align}
I &= V_2\int \of \rho\, e^{2A(\rho)}\left[\left(\kappa W(\phi)-\frac1{\kappa}\dot A(\rho)\right)^2-\frac12\left(\dot\phi(\rho)+W'(\phi)\right)^2\right] \\
&\quad+ V_2\int \of \rho\,\partial_\rho\left[e^{2A(\rho)}\left(W(\phi)-\frac2{\kappa^2}\dot A(\rho)\right)\right]  \nn ~.
\end{align}
Here $V_2=\int dtdx$. The dot denotes a derivative with respect to $\rho$, the prime a derivative with respect to $\phi$ and $W(\phi)$ is any function that solves
\be
V(\phi) = -\kappa^2W(\phi)^2+\frac12W'(\phi)^2 ~.
\ee
The function $W(\phi)$ has dimensions of length$^{-2}$. The equations of motion can now be easily obtained by setting the squares to zero,
\be
\kappa^2W(\phi) = \dot A(\rho) ~, \qquad  W'(\phi) = -\dot\phi(\rho) ~.
\ee
The perfect square structure of the action leading to these equations of motion can also be derived using the Hamilton-Jacobi formalism and is reminiscent of BPS equations \cite{deBoer:2000cz}. One can check that the equations of motion solve the full non-linear Einstein equations and the Klein-Gordon equation.

Let us now consider a holographic RG flow between two AdS spaces connected via a domain wall. From the equations of motion, we see that a stationary point $\phi_\star$ of $W(\phi)$ corresponds to a solution with AdS length $\ell = (\kappa^2W(\phi_\star))^{-1} $. Although it is not necessary for our argument to give an explicit form of the potential, we find it illustrative to work through an explicit example so we pick the simple potential
\be
W(\phi) = -\phi^4 + m\phi^2 + m^2 ~.
\ee
Here $m$ is a positive constant with dimensions of length$^{-1}$ which we choose to obey $0<m\ll 1/\kappa^2$. Focussing on the region $\phi\geq 0$, this function has two critical points,
\be
\phi_\star = 0 \qquad \text{and} \qquad \phi_\star = \sqrt{\frac{m}{2}} ~.
\ee
Solving the first-order equations of motion we obtain the following solution
\begin{align}
\phi(\rho) &=\frac{\sqrt{m}}{\sqrt{2-e^{4 m \rho-2 c_1 m}}} ~,\nn \\
A(\rho) &= \frac{1}{8} \kappa ^2 m \left(\frac{1}{e^{4 m \rho-2 c_1 m}-2}+10 m \rho\right)-\frac{1}{16} \kappa ^2 m \log \left(e^{4 m \rho}-2 e^{2 c_1 m}\right)+c_2~.
\end{align}
Here $c_1,c_2$ are integration constants. We find that the critical points of $W(\phi(r))$ are reached asymptotically,
\begin{align}
\lim_{\rho\to\infty} \phi(\rho) &= 0 ~, \\
\lim_{\rho\to-\infty} \phi(\rho) &= \sqrt{\frac{m}{2}} ~. \nn 
\end{align}
If we consider fluctuations $\delta\varphi=\phi-\phi_\star$ around the critical points, we find that $V''(\delta\varphi)<0$ around $\phi_\star=0$ and $V''(\delta\varphi)>0$ around $\phi_{\star} = \sqrt{m/2}$. Because a scalar field with mass $M$ in the bulk is dual to a CFT operator with conformal dimension $\Delta$ given by $M^2\ell^2=\Delta(\Delta-2)$, we identify $\phi_\star=0$ as the UV CFT perturbed by a relevant deformation ($\Delta < 2$) and $\phi_\star=\sqrt{m/2}$ as the IR CFT perturbed by an irrelevant deformation $(\Delta > 2)$.

The $c$-function that gives the central charge at the critical points is given by \cite{Hotta:2008xt}
\be
c(\rho) = \frac{3}{2G_3\dot A} ~.
\ee
At the critical points, we find
\begin{align}
\lim_{\rho\to\infty}A(\rho) &= \kappa^2m^2\rho~, \\
\lim_{\rho\to-\infty}A(\rho) &= \frac54\kappa^2m^2\rho \nn ~.
\end{align}
We see that the IR central charge is smaller than the UV central charge in accord with the $c$-theorem. More generally, the change in the $c$-function along the radial direction is
\be
\dot c(\rho) = -\frac{3\ddot A}{2G_3\dot A^2} ~.
\ee
The requirement that the central charge monotonically decreases (or is uncorrected) along the flow from UV to IR is $\ddot A \leq 0$. From Einstein's equations, this condition follows naturally. Contracting the Einstein tensor with a null vector $\zeta^a =(\sqrt{-g^{tt}},\sqrt{g^{\rho\rho}},0) $ we find
\be
G_{ab}\zeta^a\zeta^b = -\ddot A(\rho) ~.
\ee
The condition $\ddot A \leq0$ follows directly from the NEC ($T_{ab}\zeta^a\zeta^b\geq0$). So, as long as the NEC is obeyed, the central charge of the IR theory is smaller than the central charge of the unperturbed UV theory, so that the spinning WGC for BTZ black holes is satisfied. This is the bulk dual of the $c$-theorem in the CFT.

\section{Five-dimensional black string}\label{sec:higherdim}
Now that we have seen that perturbing a BTZ black hole by a relevant deformation leads to a spinning version of the WGC, one might wonder whether the spinning WGC has any application in constraining the extremality bound of charged extremal solutions that have near-horizon limits with BTZ factors (the particular example that we consider is a boosted five-dimensional black string). This idea is quite natural as it is well known that the entropy of such charged extremal solutions can be easily determined using Cardy's formula in the near-horizon BTZ geometry (see \cite{Sen:2007qy} for example). Because of the close connection between entropy and extremality \cite{Hamada:2018dde,Goon:2019faz} one might naively think that both extremality bounds should coincide and that the spinning WGC implies the charged WGC. One should keep in mind however that this relation only holds for the microcanonical entropy (which is not evaluated at zero temperature). 

By computing higher-derivative corrections to a five-dimensional boosted black string, we show that while the entropy of the BTZ and four-dimensional black hole agree at zero temperature, their extremality bounds do not. To determine the corrections to the near-horizon BTZ$\times S^2$ geometry we find it convenient to use $c$-extremization again and corrections to the four-dimensional black hole are computed employing a thermodynamic approach. As an additional check of our results, we also calculate corrections to the four-dimensional extremality bound using \eqref{eq:WGCcondition} and on top of that explicitly construct the five-dimensional higher-derivative corrected black string. In App. \ref{app:explicit5Dstring} we show that the explicit solution is in perfect agreement with our results from $c$-extremization, our integrated condition, and the thermodynamic approach.

\begin{figure}[t]
    \centering
    \begin{tikzpicture}[baseline={(current bounding box.center)}]
        \draw[thick] (4,0) -- (0,0);
        \draw[thick] (-1,-1) -- (3,-1);
        \draw[thick] (4,4) -- (0,4);
        \draw[thick] (-1,3) -- (3,3);
        \fill[opacity=0.2] (0,0) -- (0,4) -- (-1,3) -- (-1,-1);
        \fill[red,opacity=0.2] (1.5,0) -- (1.5,4) -- (0.5,3) -- (0.5,-1);
        \fill[red,opacity=0.2] (0.75,0) -- (0.75,4) -- (-0.25,3) -- (-0.25,-1);
        \begin{scope}[decoration={
            markings,
            mark=at position 0.5 with {\arrow{>>}}}
            ]
            \draw[thick,dashed,postaction={decorate}] (0,0) -- (0,4);
            \draw[thick,dashed,postaction={decorate}] (-1,-1) -- (-1,3);
            \draw[red,thick,dashed,postaction={decorate}] (1.5,0) -- (1.5,4);
            \draw[red,thick,dashed,postaction={decorate}] (0.5,-1) -- (0.5,3);
            \draw[red,thick,dashed,postaction={decorate}] (0.75,0) -- (0.75,4);
            \draw[red,thick,dashed,postaction={decorate}] (-0.25,-1) -- (-0.25,3);
        \end{scope}
        \draw (3.5,4.4) circle (0.4);
        \draw (2.5,4.3) circle (0.3);
        \draw[fill=red, fill opacity=0.2] (1.5,4.2) circle (0.2);
        \draw[fill=red, fill opacity=0.2] (0.75,4.1) circle (0.1);
        \draw (3.1,4.4) to[out=-25,in=180+25] (3.9,4.4);
        \draw[dotted] (3.1,4.4) to[out=25,in=180-25] (3.9,4.4);
        \draw (2.2,4.3) to[out=-25,in=180+25] (2.8,4.3);
        \draw[dotted] (2.2,4.3) to[out=25,in=180-25] (2.8,4.3);
        \draw (1.3,4.2) to[out=-25,in=180+25] (1.7,4.2);
        \draw[dotted] (1.3,4.2) to[out=25,in=180-25] (1.7,4.2);
        \draw (0.65,4.1) to[out=-25,in=180+25] (0.85,4.1);
        \draw[dotted] (0.65,4.1) to[out=25,in=180-25] (0.85,4.1);
        \draw[->] (-0.5,-1.3) --node[below] {$r$} (1.5,-1.3);
        \draw[->] (-1.3,-0.5) --node[left] {$t$} (-1.3,1.5);
        \draw[->] (-1,3.2) --node[left] {$x$} (-0.2,4);
        \node at (4.1,4.8) {$S^2$};
    \end{tikzpicture} \hspace{15pt}
    \begin{tikzpicture}[baseline={(current bounding box.center)}]
        \draw[thick] (4,2) -- (0,2);
        \draw[thick] (4,4) -- (0,4);
        \draw[thick] (4,3) ellipse (0.3 and 1);
        \draw[thick,gray] (0,2) arc (270:90:0.3 and 1);
        \draw[thick,gray,dotted] (0,2) arc (-90:90:0.3 and 1);
        \draw[thick,red] (1.5,2) arc (270:90:0.3 and 1);
        \draw[thick,red,dotted] (1.5,2) arc (-90:90:0.3 and 1);
        \draw[thick,red] (0.75,2) arc (270:90:0.3 and 1);
        \draw[thick,red,dotted] (0.75,2) arc (-90:90:0.3 and 1);
        \draw (3.5,4.4) circle (0.4);
        \draw (2.5,4.3) circle (0.3);
        \draw[fill=red, fill opacity=0.2] (1.5,4.2) circle (0.2);
        \draw[fill=red, fill opacity=0.2] (0.75,4.1) circle (0.1);
        \draw (3.1,4.4) to[out=-25,in=180+25] (3.9,4.4);
        \draw[dotted] (3.1,4.4) to[out=25,in=180-25] (3.9,4.4);
        \draw (2.2,4.3) to[out=-25,in=180+25] (2.8,4.3);
        \draw[dotted] (2.2,4.3) to[out=25,in=180-25] (2.8,4.3);
        \draw (1.3,4.2) to[out=-25,in=180+25] (1.7,4.2);
        \draw[dotted] (1.3,4.2) to[out=25,in=180-25] (1.7,4.2);
        \draw (0.65,4.1) to[out=-25,in=180+25] (0.85,4.1);
        \draw[dotted] (0.65,4.1) to[out=25,in=180-25] (0.85,4.1);
        \draw[->] (0.5,1.7) --node[below] {$r$} (2.5,1.7);
        \draw[->] (-0.4,2.2) to[out=105,in=255] node[left] {$x$} (-0.4,3.8);
        \node at (4.1,4.8) {$S^2$};
    \end{tikzpicture}
    \caption{Sketch of the five-dimensional black string geometry. The coordinate $x$ is periodic, shown with identifications on the left, while on the right $t$ is suppressed. In the coordinates of \eqref{eq:5Dboosted} the ring singularity is at $r=-q$ and the two horizons (red), located at $r=0$ and $r=r_0$, have topology $S^1\times S^2$ at fixed $t$. The four-dimensional black hole is found by reducing over $x$ (leading to a horizon of spacial topology $S^2$ and a point-like singularity). The $\text{BTZ}\times S^2$ geometry is found in the near-horizon limit.}
    \label{fig:5Dgeometry}
\end{figure}
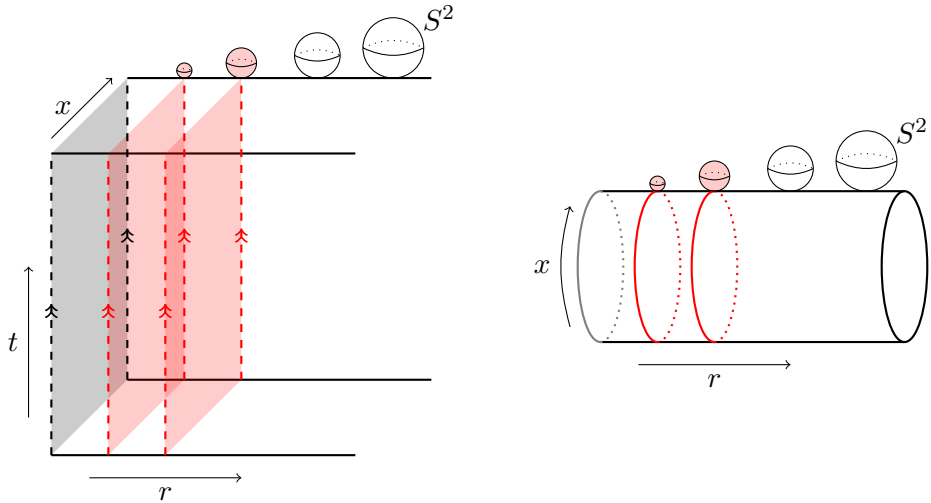

The precise system we will study is a five-dimensional boosted black string that is a solution to the Einstein-Maxwell action perturbed by higher-derivative terms.\footnote{Such a geometry can arise for example by compactifying three (magnetically) charged intersecting M5-branes, see \cite{Balasubramanian:1998ee}.}
Before adding higher-derivative corrections the relevant action is given by
\be
I = \frac1{16\pi G_5}\int \mathrm{d}^5x\,\sqrt{-g}\left(R-\frac34F_{MN}F^{MN}\right) ~.
\ee
The line element that describes an unboosted black string extended along a compact $x$-direction is given by \cite{Balasubramanian:1998ee}
\be
\of s^2 = H(r)^{-1}\left[-\left(1-\frac{r_0}{r}\right)\of t^2+\of x^2\right] + H(r)^{2}\left[\left(1-\frac{r_0}{r}\right)^{-1}\of r^2+r^2\of \Omega_2^2\right] ~.
\ee
The factor of three in front of the Maxwell term in the action signifies that we consider a setup with three (equal) magnetic charges. The magnetic field strength $F$ and harmonic function $H(r)$ are defined as
\begin{align}
F &= \sqrt{q(q+r_0)}\,\sin\theta\, \of \theta\wedge\of\phi ~, \\
H(r) &= 1+\frac qr~. \nn
\end{align}
The physical charge of the solution is given by
\be
Q = \frac1{4\pi}\int_{S^2_\infty} F = \sqrt{q(q+r_0)} ~.
\ee
We can now perform a boost along the $x$-direction, which transforms
\begin{align}
t &\to \cosh\delta_0 \,t + \sinh\delta_0 \,x \,,\\
x &\to \sinh\delta_0 \,t + \cosh\delta_0 \,x ~.
\end{align}
The metric now becomes
\begin{align} \label{eq:5Dboosted}
\of s^2 &=  H(r)^{-1}\left[-\of t^2 +\of x^2 +\frac{r_0}{r}\left(\cosh\delta_0 \,\of t+\sinh\delta_0 \,\of x\right)^2\right] \\
&\quad+H(r)^{2}\left[\left(1-\frac{r_0}{r}\right)^{-1}\of r^2 + r^2\of \Omega_2^2\right] ~. \nn
\end{align}
As we will now show, this geometry describes a four-dimensional charged black hole after a Kaluza-Klein reduction and has a BTZ$\times S^2$ near-horizon limit: see Fig.~\ref{fig:5Dgeometry} for a sketch.

\subsection{Four-dimensional black hole}
To obtain a charged four-dimensional black hole solution, we take $x=x+2\pi R$ to be compact and perform a Kaluza-Klein reduction. We take the standard ansatz
\be \label{eq:KKansatz}
\of s^2 = g_{\mu\nu}\of x^\mu \of x^\nu + \varphi^2\big[(A_0)_\mu\of x^\mu + \of x\big]^2 ~,
\ee
where $\varphi$ is a scalar field and $A_0$ a Kaluza-Klein gauge field corresponding to the electric field $F_0=\of A_0$. Defining $q_0=r_0\sinh^2\delta_0$ we read off
\be
\varphi^2 = H(r)^{-1}H_0(r) ~, \quad A_0 = \frac{\sqrt{q_0(q_0+r_0)}}{q_0+r}\,\of t ~,
\ee
with
\be
H_0(r) = 1+\frac{q_0}r ~.
\ee
After performing the reduction and going to Einstein frame, we find that the metric \eqref{eq:5Dboosted} becomes the following four-dimensional black hole,
\be \label{eq:4Dblackhole}
\of s^2 = - f(r) \of t^2 + f(r)^{-1} \of r^2 + H_0(r)^{1/2}H(r)^{3/2}r^2\,\of \Omega_2^2 ~,
\ee
with
\be
f(r) = H_0(r)^{-1/2}H(r)^{-3/2}\left(1-\frac{r_0}{r}\right) ~.
\ee
It will be useful to combine the electric field arising from the Kaluza-Klein reduction and magnetic field already present in five dimensions into a single field strength defined by
\be
{\cal F} = \frac{\sqrt{q_0(q_0+r_0)}}{r^2}H(r)^{-2}\,\of t\wedge \of r + \sqrt{3q(q+r_0)}\,\sin{\theta}\,\of\theta\wedge\of\phi ~. 
\ee
The physical electric and three magnetic charge of the black holes are given by
\be
Q_0 = \frac1{4\pi}\int_{S^2_\infty} \star F_0 = \sqrt{q_0(q_0+r_0)} ~, \quad Q = \frac1{4\pi}\int_{S^2_\infty} F = \sqrt{q(q+r_0)} ~.
\ee
For simplicity, we now also set the electric charge equal to the magnetic charge, i.e. $q_0=q$, such that $\varphi=1$. The four-dimensional action then takes the standard form
\be
I = \frac1{16\pi G_4}\int \of^4x\sqrt{-g}\left(R - \frac14 {\cal F}_{\mu\nu}{\cal F}^{\mu\nu}\right) ~.
\ee
Having obtained the black hole solution, it is easy to find the extremality bound and entropy of this black hole. Solving $f(r)=0$ we find that the horizons are located at
\be
r_+= r_0 ~, \quad r_-=0 ~,
\ee
and the extremal limit corresponds to $r_0\to0$. From the asymptotic form of $f(r)$
\be
\lim_{r\to\infty}f(r) =  1 - \frac{2q+r_0}{r} + {\cal O}\Big(\frac{1}{r^2}\Big) ~,
\ee
we find that the ADM mass is given by
\be
2G_4M_4 = 2q+r_0 ~.
\ee
In terms of the physical charges, the extremality bound is given by
\be
\frac{Q}{G_4M_4} = \frac{2\sqrt{q(q+r_0)}}{2q+r_0} \leq 1 ~.
\ee
Finally, the Bekenstein-Hawking entropy of the extremal black hole is given by
\be
S\big|_{T=0} = \frac{A}{4G_4} = \frac{\pi Q^2}{G_4} ~.
\ee

\subsection{\texorpdfstring{$\mathrm{BTZ}\times S^2$}{BTZxS2} solution}
We now consider the near-horizon limit of the boosted black string solution \eqref{eq:5Dboosted} by taking the limit $q\gg r$, which sends
\be
H(r) \to \frac{q}{r} ~.
\ee
After performing the coordinate transformation
\be
\tau = \frac{t\ell}{R}~, \qquad \rho^2 = \frac{2R^2}{\ell}\left(r+q\right) ~, \qquad \psi  =\frac{x}{R} ~,
\ee
we obtain a $\text{BTZ}\times S^2$ solution.
\be
\of s^2 = -N(\rho)^2\,\of \tau^2 + N(\rho)^{-2}\,\of\rho^2 +\rho^2\big(\of \psi+N^\psi(\rho) \,\of\tau\big)^2 + q^2\,\d{\Omega_2^2} ~.
\ee
Explicitly, the lapse and shift functions are given by
\begin{align}
N(r)^2 &= \frac{\left(\rho^2 - R^2\right) \big(\rho^2-(1+\frac{r_0}{q})R^2\big)}{4 q^2 \rho ^2} ~, \\
N^\psi(\rho) &= \frac{R^2}{2\rho^2}\sqrt{\frac{q+r_0}{q^3}}\nn ~.
\end{align}
The BTZ black hole has an AdS length $\ell = 2q$ and the $S^2$ has radius $\ell_{S^2}=q$. The mass and angular momentum are given by
\be \label{eq:3D4Drelation}
M_3 = \frac{R^2(\ell+r_0)}{4G_3\ell^3} ~, \quad J_3 = \frac{R^2}{4G_3}\sqrt{\frac{\ell+2r_0}{\ell^3}} ~.
\ee
The extremal limit corresponds to $r_0=0$ and the extremality bound is given by
\be
\frac{J_3}{\ell M_3} = \frac{\sqrt{\ell(\ell+2r_0)}}{\ell+r_0} \leq 1 ~.
\ee
Instead of explicitly performing the near-horizon limit an alternative way of obtaining the same $\text{BTZ}\times S^2$ solution is to use $c$-extremization \cite{Kraus:2005vz}. This will be especially useful when we include higher-derivative corrections. Going to Euclidean signature and evaluating the five-dimensional action, we find that the $c$-function takes the form
\be
c(\ell,\ell_{S^2}) = \frac{3\pi}{2G_5}\ell^2\ell_{S^2}^2\left(-R+\frac34F_{MN}F^{MN}\right) = \frac{3 \pi  \ell \left[\ell^2 \left(3 Q^2-4 \ell_{S^2}^2\right)+12 \ell_{S^2}^4\right]}{4 G_5 \ell_{S^2}^2} ~.
\ee
Extremizing with respect to $\ell$ and $\ell_{S^2}$ we find the following lengths and central charge
\begin{align}
\ell = 2Q ~, \qquad \ell_{S^2} = Q ~, \qquad c = \frac{3Q}{G_3} ~,
\end{align}
where we used $G_5 = 4\pi\ell_{S^2}^2G_3$. This coincides with the solution found by taking the near-horizon limit. The entropy can be found using Cardy's formula \cite{Kraus:2006wn}
\be
S = 2\pi\sqrt{\frac{c}{6}\left(h-\frac c{24}\right)}+2\pi\sqrt{\frac{c}{6}\left(\bar h-\frac c{24}\right)} ~.
\ee
After using \eqref{eq:MJh}, we find that in the extremal limit
\be
S\big|_{T=0} = \pi\sqrt{\frac{\ell^2M_3}{G_3}} = \frac{\pi Q^2}{G_4} ~,
\ee
where we wrote $G_3 = \frac{RG_4}{2Q^2}$. The extremal entropy of the BTZ matches that of the extremal four-dimensional black hole.

\subsection{Including higher-derivative corrections}
\label{sec:hdcorrectedblackstring}
We are now ready to include higher-derivative corrections to the five-dimensional action and see how the BTZ near-horizon geometry and four-dimensional black hole are modified. The most general four-derivative corrections to five-dimensional Einstein-Maxwell theory are
\begin{align}\label{eq:5Daction}
I = \frac1{16\pi G_5}\int \of^5x\,\sqrt{-g}\,&\Big(R-\frac34F_{MN}F^{MN} + \alpha_1Q^2F_{MN}F^{MN}F_{OP}F^{OP} \\
& {}+ \alpha_2 Q^2 F_{MN}F_{OP}W^{MNOP} + \alpha_3 Q^2 E_5 \Big) ~. \nn
\end{align}
Here $W_{MNOP}$ is the Weyl tensor and 
\be
E_5 = R_{MNOP}R^{MNOP}-4R_{MN}R^{MN}+R^2~,
\ee
is the five-dimensional Euler density. We normalize the higher-derivative operators with $Q$ so that the $\alpha_i$ are dimensionless.

Readers uninterested in technical details can skip the next two subsections and instead directly look at Table \ref{tab:summary}, where we give an overview of the corrections to the entropy and extremality bounds. Take notice that the BTZ extremality bound does not coincide with the four-dimensional extremality bound.

\begin{table}[t]
    \centering
    \renewcommand*{\arraystretch}{1.75}
    \newcolumntype{R}{>{$}r<{$}}
    \newcolumntype{L}{>{$}l<{$}}
    \newcolumntype{M}{R@{${}={}$}L}
    \begin{tabular}{|c|c|ML|}
        \hline
        \parbox[t]{3mm}{\multirow{4}{*}{\rotatebox[origin=c]{90}{$\text{BTZ}\times S^2$}}} & \multirow{2}{*}{$T=0$} & z & 1 + \frac{8\alpha_1+3\alpha_2-12\alpha_3}{2} & \\
         & & S & 2\pi Q\sqrt{\frac{M_3}{G_3}}\left(1 - \frac{8\alpha_1+3\alpha_2-12\alpha_3}{2}\right) & = \frac{\pi Q^2}{G_4}\left(1 - \frac{8\alpha_1+3\alpha_2-12\alpha_3}{2}\right)\\ \cline{2-5}
         & \multirow{2}{*}{$z=1$} & T & \sqrt{\frac{G_3J_3(8\alpha_1+3\alpha_2-12\alpha_3)}{\pi Q^3}} & \\
         & & S & 2\pi Q\sqrt{\frac{M_3}{G_3}}\Big(1 + \sqrt{\frac{8\alpha_1+3\alpha_2-12\alpha_3}{2}}\Big) & = \frac{\pi Q^2}{G_4}\Big(1 +  \sqrt{\frac{8\alpha_1+3\alpha_2-12\alpha_3}{2}}\Big)\\[3pt] \hline\hline
        \parbox[t]{3mm}{\multirow{4}{*}{\rotatebox[origin=c]{90}{4D}}} & \multirow{2}{*}{$T=0$} & z & 1 + \frac{2a_1+a_2}{10} & = 1 + \frac{8\alpha_1+7\alpha_2+6\alpha_3}{40}\\
         & & S & \frac{\pi Q^2}{G_4}\left(1 - 4a_1+4a_3\right) & = \frac{\pi Q^2}{G_4}\left( 1 - \frac{8\alpha_1+3\alpha_2-12\alpha_3}{2} \right)\\ \cline{2-5}
         & \multirow{2}{*}{$z=1$} & T & \frac{\pi}{Q}\sqrt{\frac{2(2a_1+a_2)}{5}} & = \frac{\pi}{Q}\sqrt{\frac{8\alpha_1+7\alpha_2+6\alpha_3}{10}}\\
         & & S & \frac{\pi Q^2}{G_4}\Big(1 + \sqrt{\frac{2(2a_1+a_2)}{5}}\Big) & = \frac{\pi Q^2}{G_4}\Big(1 + \sqrt{\frac{8\alpha_1+7\alpha_2+6\alpha_3}{10}}\Big)\\[3pt] \hline
    \end{tabular}
    \caption{Overview of the corrections to the extremality bounds, entropies and temperatures. $z=\frac{J_3}{2QM_3}$ for the BTZ black hole and $z=\frac{Q}{G_4M_4}$ for the four-dimensional black hole. It is clear that thermodynamics at $z=1$ only makes sense if the WGC is satisfied (else $z=1$ is a naked singularity with no horizon). Results for $\text{BTZ}\times S^2$ are presented also in terms of ``four-dimensional quantities'' via the correspondence $G_3=\frac{2\pi R}{4\pi Q^2} G_4$. The relation between $a_i$s and $\alpha_i$s are given in \eqref{eq:Romancoefficients}.}
    \label{tab:summary}
\end{table}

\subsubsection{Four-dimensional black hole solution}
To obtain the four-dimensional action with higher-derivative corrections we perform a Kaluza-Klein reduction along the $x$-direction, just as before, using the ansatz \eqref{eq:KKansatz}. Taking magnetic and electric charges equal the reduced action takes the following form.
\begin{align}\label{eq:4Dhd}
I = \frac1{16\pi G_4}\int \d[4]{x}\,\sqrt{-g}\,&\Big(R-\frac14{\cal F}_{\mu\nu}{\cal F}^{\mu\nu} + \frac{a_1}{4}Q^2\big({\cal F}_{\mu\nu}{\cal F}^{\mu\nu}\big)^2\\
&\qquad{}+ \frac{a_2}{2} Q^2{\cal F}_{\mu\nu}{\cal F}_{\rho\sigma}W^{\mu\nu\rho\sigma} + \frac{a_3}{2} Q^2E_4 \Big) ~. \nn
\end{align}
The Wilson coefficients are related to the coefficients appearing in the five-dimensional action as
\be \label{eq:Romancoefficients}
a_1 = \alpha_1  +\frac38\alpha_2 + \frac12\alpha_3 ~, \qquad a_2 = \alpha_2+\frac12\alpha_3 ~, \qquad a_3 = 2\alpha_3 ~.
\ee
Because the four-dimensional Euler density $E_4$ is topological, it will affect neither the equations of motion nor the extremality bound. To determine the corrections to the extremality bound and entropy, we will make use of a thermodynamic approach, which has the advantage that we don't need to explicitly know the corrected metric. Instead, we can evaluate the corrected Euclidean action on the uncorrected solution \cite{Reall:2019sah,Loges:2019jzs}. As an additional check, we show in App. \ref{app:explicit5Dstring} that this approach agrees with a direct computation of the corrected metric.

The Euclidean action is given by
\begin{align}
I_E =& -\frac1{16\pi G_4}\int \d[4]{x} \,\sqrt{-g}\,\Big(R-\frac14{\cal F}_{\mu\nu}{\cal F}^{\mu\nu} + \frac{a_1}{4}Q^2{\cal F}_{\mu\nu}{\cal F}^{\mu\nu}{\cal F}_{\rho\sigma}{\cal F}^{\rho\sigma} \\
&\qquad {}+ \frac{a_2}{2} Q^2{\cal F}_{\mu\nu}{\cal F}_{\rho\sigma}W^{\mu\nu\rho\sigma} + \frac{a_3}{2} Q^2E_4 \Big) - \frac1{8\pi G_4}\oint d^3x\sqrt{h}\left(K-K_0\right)\nn ~,
\end{align}
where we supplemented the bulk action by a Gibbons-Hawking-York boundary term. Here $K$ is the extrinsic curvature of the induced metric at the boundary $r\to\infty$ and $K_0$ a counterterm constructed by embedding the boundary metric in flat space. This is required to obtain a finite on-shell action.

Before we take into account the higher-derivative corrections, we first focus on the uncorrected action. By explicitly evaluating the uncorrected Euclidean action on the uncorrected black hole background \eqref{eq:4Dblackhole} one finds
\be
I_E = \frac{\beta}{2}\left(M_4-\Phi \tilde Q_0 + 3\Psi \tilde Q\right) ~.
\ee
The factor of three indicates there are three magnetic charges. Here $\beta=T^{-1}$ is the inverse temperature, $\tilde Q = Q/4G_4$ and $\tilde Q_0=Q_0/4G_4$ are rescaled charges and $\Phi$ and $\Psi$ are the electric and magnetic potentials given by
\be
\Phi = \frac{q_0}{\sqrt{q_0(q_0+r_0)}} ~, \quad \Psi = \frac{q}{\sqrt{q(q+r_0)}} ~.
\ee
Using the Smarr formula $M_4=2TS+\Phi\tilde Q_0+3\Psi \tilde Q$, the Euclidean action can now be written as
\be
I_E = \beta G = \beta(M_4-TS-\Phi \tilde Q_0) ~,
\ee
with $G$ the Gibbs free energy. This relation still holds in the presence of higher-derivative corrections as long as we interpret $S$ as the Wald entropy \cite{Reall:2019sah}.

To derive the various thermodynamic quantities of the black hole, we express the Gibbs free energy as $G=G(T,\Phi,\tilde Q)$, appropriate for a grand canonical ensemble. The entropy and mass of the black hole are given by
\be
S=-\left(\frac{\partial G}{\partial T}\right)_{\Phi,\tilde Q} ~.
\ee
After having obtained the entropy, the mass is given by
\be
M_4 = G +\Phi\tilde Q_0 + TS ~.
\ee
We now include higher-derivative corrections. Evaluating the corrected Euclidean action on the uncorrected black hole background and using the thermodynamic identities, we find that the $T=0$ entropy is given by
\be \label{eq:4DentropyZeroTemp}
S\big|_{T=0} =\frac{\pi Q^2}{G_4}\left(1-4a_1+4a_3\right) ~,
\ee
where we took equal electric and magnetic charges. At zero temperature the correction to the mass is
\be
G_4M_4 = Q\left( 1 - \frac{2a_1+a_2}{10}\right)  ~.
\ee
The extremality bound is therefore modified as
\be \label{eq:chargedextremality}
z\equiv \frac{Q}{G_4M_4} \leq 1 + \frac{2a_1+a_2}{10} ~.
\ee
Finally, we are also interested in the microcanonical entropy of the $z=1$ black hole in the corrected theory. Expanding the mass at small temperature we find that this black hole has a non-zero temperature
\be \label{eq:finitetemp}
T\big|_{z=1} = \frac{\pi}{Q}\sqrt{\frac{2(2a_1+a_2)}{5}} ~.
\ee
The entropy becomes
\be \label{eq:4Dentropy}
S\big|_{z=1} = \frac{\pi Q^2}{G_4}\left(1+ \sqrt{\frac{2(2a_1+a_2)}{5}}\right) ~.
\ee
The WGC is satisfied when the extremality bound is corrected positively, which requires
\be\label{eq:4Dbound}
2a_1 + a_2 \geq 0 \quad \leftrightarrow \quad 8\alpha_1 + 7\alpha_2 + 6\alpha_3 \geq 0 ~.
\ee
The same bound can be derived using \eqref{eq:WGCcondition}. Because the Euler density does not contribute to the extremality bound in four dimensions, we can use the previously derived stress tensor and corrections to Maxwell's equations (see \eqref{eq:4DStressMod} and \eqref{eq:MaxwellMod}, but note the different normalization of the action \eqref{eq:4Dhd}). We then find
\be
\int_\Sigma \mathrm{d}^{3}x \,\sqrt{h}\,\delta T^{\rm eff}_{ab}\xi^an^b = -\frac{Q}{10 G_4}\left(2 a_1+a_2\right) ~.
\ee
Imposing the integrated stress tensor to be smaller than or equal to zero, we again find $2a_1+a_2\geq0$.

\subsubsection{\texorpdfstring{$\mathrm{BTZ}\times S^2$}{BTZxS2} solution}
To find the $\text{BTZ}\times S^2$ near-horizon geometry in the corrected theory, it is very convenient to use $c$-extremization instead of computing the corrected metric. Taking the same ansatz for the metric as before, we find that the $c$-function takes the form
\begin{align}
c(\ell,\ell_{S^2}) 
&= \frac{3 \pi  \ell \left[\ell^2 \left(3 Q^2-4 \ell_{S^2}^2\right)+12 \ell_{S^2}^4\right]}{4 G_5 \ell_{S^2}^2} \\
&\qquad+ \frac{3 \pi  \ell Q^2 \left[-2 \alpha _1 \ell^2 Q^4+\alpha _2 Q^2 \left(\ell_{S^2}^2-\ell^2\right) \ell_{S^2}^2+12 \alpha _3 \ell_{S^2}^6\right]}{G_5 \ell_{S^2}^6} ~. \nn
\end{align}
Extremizing the $c$-function for $\ell$ and $\ell_{S^2}$ we find the following lengths:
\begin{align}
\ell &= 2 Q - \frac{2Q}{3}\left(\alpha_2-3\alpha_3\right) ~,\\
\ell_{S^2} &= Q-\frac{Q}{4}\left(8\alpha_1+3\alpha_2-2\alpha_3\right)\nn ~.
\end{align}
Because the $S^2$ length is also corrected by higher-derivative terms, this modifies the relationship between the five-dimensional and three-dimensional Newton constant. We are interested in computing corrections keeping the Newton constant fixed, so we either have to express corrections to the central charge in terms of $G_5$ or rescale $G_3$ to keep the ratio $G_5/G_3$ uncorrected. We choose the latter option and therefore rescale
\be
G_3 \to G_3 Q^2/\ell_{S^2}^2,
\ee
and express all correction to the BTZ geometry with respect to this rescaled Newton constant. The relationship between the different Newton constants is then still given by
\be
G_5 = 4\pi Q^2G_3 = 2\pi R G_4 ~.
\ee
Using the values for the AdS and $S^2$ length we found the central charge is given by
\be
c = \frac{3Q}{G_3}\left(1-\frac{8\alpha_1+3\alpha_2-12\alpha_3}{2}\right) ~.
\ee
To compute the entropry, we use Cardy's formula and find that the entropy at zero temperature is given by
\be
S\big|_{T=0} = 2\pi Q\sqrt{\frac{M_3}{G_3}}\left(1-\frac{8\alpha_1+3\alpha_2-12\alpha_3}{2}\right) =\frac{\pi Q^2}{G_4}\left(1-\frac{8\alpha_1+3\alpha_2-12\alpha_3}{2}\right) ~.
\ee
Using \eqref{eq:Romancoefficients} this entropy equals \eqref{eq:4DentropyZeroTemp}. 

To derive the correction to the extremality bound in a canonical ensemble and the entropy in a microcanonical ensemble we again use a thermodynamic approach, just as in Sec. \ref{sec:holoRG}. The five-dimensional Euclidean action supplemented by a Gibbons-Hawking-York boundary term and counterterm is given by
\begin{align}
I_E &= \frac1{16\pi G_5}\int \of^5x \,\sqrt{g}\,\Big(-R+\frac34F_{MN}F^{MN} - \alpha_1Q^2F_{MN}F^{MN}F_{OP}F^{OP} \\
& \qquad\qquad {} - \alpha_2 Q^2 F_{MN}F_{OP}W^{MNOP} - \alpha_3 Q^2 E_5 \Big) -\frac1{8\pi G_5}\oint d^4x\sqrt{h}\left(K-K_0\right) ~. \nn
\end{align}
We now evaluate this on a $\text{BTZ}\times S^2$ background. To make the on-shell action finite, the counterterm is chosen to be
\be
K_0 = \frac1{2Q}\left(1+\frac{8\alpha_1+3\alpha_2-12\alpha_3}{2}\right) ~.
\ee
Using this the complete on-shell action evaluates to
\be
I_E = \frac{\pi  \beta  \left(r_-^2-r_+^2\right)}{8 G_5}-\frac{\pi \beta \left(8 \alpha _1+3 \alpha _2-12 \alpha _3\right)\left(r_-^2-r_+^2\right)}{16 G_5} ~.
\ee
Just as before, we can write
\be
I_E = \beta G = \beta\left(M_3-TS-\Omega J_3\right) ~,
\ee
and the thermodynamic quantities are given by \eqref{eq:3Dthermo} and \eqref{eq:3Dmass}. We find that the extremality bound in a canonical ensemble is corrected as
\be
\frac{J_3}{2QM_3} \leq 1  + \frac{8\alpha_1+3\alpha_2-12\alpha_3}{2} ~,
\ee
so that this spinning WGC is satisfied when
\begin{equation}\label{eq:BTZbound}
    8\alpha_1+3\alpha_2-12\alpha_3 \geq 0
\end{equation}
Notably, this combination of Wilson coefficients does not coincide with the combination appearing in the extremality bound of the four-dimensional black hole. The state $z=J_3/(2QM_3)=1$ has a temperature
\be
\left. T\right|_{z=1} = \sqrt{\frac{G_3J_3(8\alpha_1+3\alpha_2-12\alpha_3)}{\pi Q^3}} ~.
\ee
At this temperature, the microcanonical entropy is given by
\be
\left.S\right|_{z=1} = 2\pi Q\sqrt{\frac{M_3}{G_3}}\left(1+\sqrt{\frac{8\alpha_1+3\alpha_2-12\alpha_3}{2}}\right) ~.
\ee
A summary of all corrections to the extremality bounds and the entropy are displayed in Table \ref{tab:summary}.

\subsection{WGC bounds}

In Sec.\ \ref{sec:hdcorrectedblackstring} we demonstrated that the WGC as phrased in terms of a corrected extremality bound differs for two distinct limits of the five-dimensional black string. The corrected angular momentum-to-mass ratio of the near-horizon $\text{BTZ}\times S^2$ black hole and corrected charge-to-mass ratio of the four-charge black hole after a Kaluza-Klein reduction to four dimensions depend on different combinations of the five-dimensional Wilson coefficients.

\begin{figure}[t]
    \centering
    \includegraphics[width=0.95\textwidth]{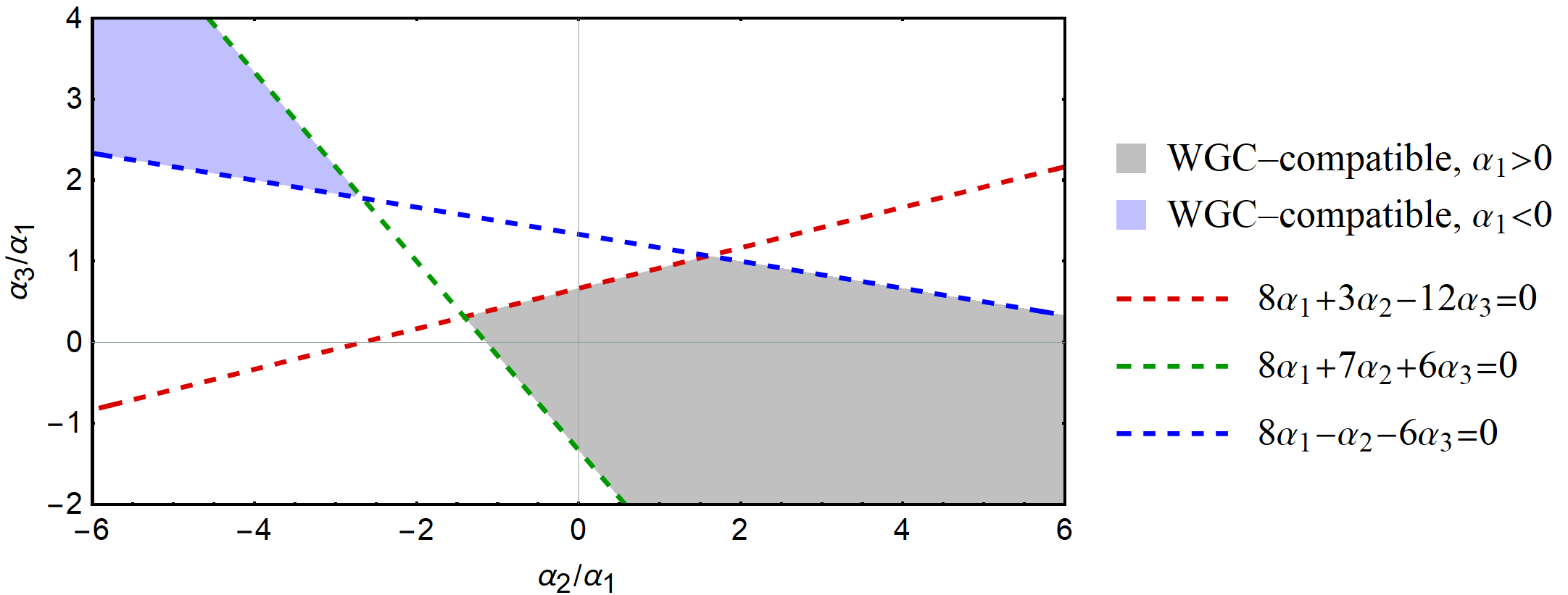}
    \caption{Comparison of the complementary bounds in Eq.~\eqref{eq:allbounds}. The dashed lines show equality and the gray and blue shaded regions show where all three \emph{in}equalities are simultaneously satisfied with $\alpha_1>0$ and $\alpha_1<0$, respectively.}
    \label{fig:regions}
\end{figure}

Alternatively, we may impose that the mild form of the WGC holds for each of these independently, allowing us to more strongly constrain the $\alpha_i$ appearing in five dimensions. This is similar in spirit to the works of~\cite{Heidenreich:2015nta,Andriolo:2018lvp} where the lattice and tower WGC respectively were argued for based on robustness under toroidal compactifications.\footnote{This of course assumes that the WGC also needs to be satisfied in the compactified theory. We refer to this as the \emph{Total Landscaping Principle}: swampland conjectures should not only be satisfied in a single theory, but also in compactifications thereof.} In fact, we can go further than only combining the bounds of Eqs.~\eqref{eq:4Dbound} and \eqref{eq:BTZbound} by asking that the mild WGC be satisfied also for electric black holes in five dimensions. Such bounds for charged black holes are known, appearing for example in~\cite{Kats:2006xp}. With the normalizations of Eq.~\eqref{eq:5Daction}, these three bounds read
\begin{equation}\label{eq:allbounds}
\begin{aligned}
    &\hspace{-90pt}\left. \begin{array}{rl}
    8\alpha_1+3\alpha_2-12\alpha_3 &\geq 0 \qquad (\text{near-horizon BTZ}\times S^2)\\[3pt]
    8\alpha_1+7\alpha_2+6\alpha_3 &\geq 0 \qquad (\text{4D 4-charge black hole})\\
    \end{array} \right\} \;\;\parbox{2cm}{\raggedright 5D boosted black string}\hspace{-27pt}\\
    8\alpha_1-\alpha_2-6\alpha_3 &\geq 0 \qquad (\text{5D electric black hole})
\end{aligned}
\end{equation}
These conditions are compatible with one another, as shown in Fig.~\ref{fig:regions}, and together provide more stringent bounds on the allowed values of the $\alpha_i$. One could also ask what bounds arise for more general charged black holes in four dimensions after a Kaluza-Klein reduction, but since both the radion and axion can be sourced we do not consider such backgrounds here.

\section{Discussion}
Understanding precisely the neccesary and sufficient conditions for proving the mild form of the WGC is an interesting question that can shed light on the boundary between those effective theories which are consistent with quantum gravity (the landscape) and those that are pathological (the swampland). In particular, one may wonder what sorts of matter configurations correct the extremality bound in a manner consistent with the WGC. To understand this better, we reformulated the shift in the extremality bound of a black hole in terms of an integrated condition on the stress tensor. When this integral of the stress tensor is negative, the horizon is shifted positively in a microcanonical ensemble. As a particular application we evaluated this condition for four-dimensional Reissner-N\"ordstrom and rotating BTZ black holes perturbed by higher-derivative corrections, but it can be applied to any stationary black hole and more general corrections. 

Applying this condition to extremal rotating BTZ black holes suggests a spinning version of the WGC that posits that corrections to the extremality bound should increase the extremal angular momentum-to-mass ratio. Although the spinning WGC does not follow from standard arguments of black hole decay, we showed that when a BTZ black hole is perturbed by a relevant operator it obeys the spinning WGC as a consequence of the $c$-theorem in the dual two-dimensional CFT.

We then studied the spinning WGC in the context of a five-dimensional boosted black string with higher-derivative corrections. The string has a near-horizon $\text{BTZ}\times S^2$ geometry and describes a four-dimensional charged black hole upon a Kaluza-Klein reduction. While the entropy of the four-dimensional black hole at zero temperature agrees with the entropy computed from the BTZ geometry, their extremality bounds do not coincide. By applying both the spinning and charged WGC to the black string we derived positivity conditions on the five-dimensional Wilson coefficients that are stronger than those obtained by applying the charged WGC alone.

Because the three-dimensional spinning WGC does not directly imply the four-dimensional charged WGC, our findings agree with the phenomenon that IR consistency is not sufficient to prove the charged WGC in $d\geq 4$. While the $c$-theorem can be used to prove the spinning WGC in three dimensions, still more UV information is needed to prove the charged WGC in higher dimensions.

In future work, it would be interesting to consider the relationship between holographic RG flow and the WGC in more detail in higher dimensions. While higher-derivative corrections have constant magnitude in a BTZ background, in higher-dimensional theories these terms vary as one moves inward from the boundary, so that perturbed geometries are more directly related to holographic RG flows. At least for a subclass of higher-derivative corrections, holographic $c$-theorems have been studied in detail \cite{Myers:2010tj}. As a particular example, we could perturb an AdS$_5$ background and use the Hamilton-Jacobi formalism to derive a function that monotonically decreases along the holographic RG flow, which would be the dual of the $a$-theorem \cite{Komargodski:2011vj} in the CFT$_4$. An important difference with three dimensions, however, is that the extremality bound is now not just determined by one anomaly coefficient; in five dimensions there are four independent four-derivative operators that contribute to an Einstein-Maxwell theory \cite{Hamada:2018dde}. Thus, to constrain the extremality bound one would have to consider a subclass of theories for which effectively only the $a$-anomaly coefficient contributes. In four-dimensional flat space, a similar strategy was employed in \cite{Charles:2019qqt} by considering the deep IR where only the $c$-anomaly contributes to the extremal charge-to-mass ratio.

Furthermore, because of the close connection between holographic $c$-theorems and entanglement entropy \cite{Myers:2010tj} it would be interesting to understand better if and how quantum corrections to the entanglement entropy are related to the WGC. In \cite{Cottrell:2016bty} for example, loop corrections to extremal black holes were studied. Those corrections modify the black hole entropy and can be viewed as a correction due to entanglement. Also, the holographic proof of the WGC presented in \cite{Montero:2018fns}, entanglement entropy played a crucial role. Because logarithmic quantum corrections to the (von Neumann) entropy of black holes are universal and determined by anomaly coefficients \cite{Solodukhin:2011gn}, one might also hope to extract similar general lessons about corrections to the extremality bound.

\section*{Acknowledgments}
We thank the authors of \cite{Cremonini:2020smy} for sharing a draft of their letter. In addition, we acknowledge useful discussions with Sera Cremonini, Brian McPeak, and Miguel Montero. This work is supported in part by the DOE under grant DE-SC0017647, the National Science Foundation under Grant No.\ NSF PHY-1748958, and the Kellett Award of the University of Wisconsin. We gratefully acknowledge the hospitality of the Kavli Institute for Theoretical Physics during the workshop ``The String Swampland and Quantum Gravity Constraints on Effective Theories'' where part of this work was completed.

\appendix

\section{Covariant phase space formalism}\label{app:covphasespace}
In this appendix we review the Iyer-Wald formalism \cite{Wald:1993nt,Iyer:1994ys} and derive some useful identities that will be used in the main body of this article. We mainly follow the notation of appendix C of \cite{Maxfield:2019hdt} and adapt that derivation to include a Maxwell term. Another more extensive review can be found here \cite{Compere:2018aar}.

\subsection{Notation and conventions}
Because this appendix heavily relies on the use of differential forms, we briefly list our conventions. A $p$-form $\alpha$ is defined as
\be
\frac1{p!}\alpha_{a_1\dots a_p}\of x^{a_1}\wedge\cdots\wedge \of x^{a_{p}} ~.
\ee 
For integration over a $d$-dimensional space, we use the following volume form
\be
\epsilon = \frac1{d!}\epsilon_{a_1\dots a_d}\of x^{a_1}\wedge\cdots\wedge \of x^{a_{d}}=  \sqrt{|g|}\;\of x^1\wedge \cdots \wedge \of x^d ~,
\ee
where $\epsilon^{12\dots d} = 1$ denotes the Levi-Civita symbol. The Hodge star operator acts on $p$-forms as
\be
{\star \alpha}  = \frac{\sqrt{|g|}}{p!(d-p)!}\,\alpha_{a_1\dots a_p}\epsilon^{a_1\dots a_p}\epsilon_{b_1\dots b_{d-p}}\,\of x^{b_1}\wedge\cdots\wedge \of x^{b_{d-p}} ~.
\ee
The Levi-Civita symbol is $\epsilon^{12\dots d} = 1$. The exterior derivative acts as
\be
\of\alpha = \left(\partial_a\omega_{b_1\dots b_p}\right)\frac1{p!}\,\of x^a\wedge \of x^{b_1}\wedge\cdots\wedge \of x^{b_{p}}~,
\ee
and obeys
\be
\of(\alpha\wedge \beta) = \of\alpha \wedge \beta +(-1)^p \alpha\wedge \of\beta ~.
\ee
Taking a $p$-form $\alpha$ and a $q$-form $\beta$, we can write
\be
\alpha\wedge \star \beta = \alpha_{a_1\dots a_p}\beta_{b_1\dots b_q} \frac{\sqrt{|g|}}{p!q!(d-q)!} \epsilon^{b_1\dots b_q}\epsilon_{c_1\dots c_{d-q}} \of x^{a_1}\wedge\dots\wedge \of x^{a_p}\wedge \of x^{c_1}\wedge\dots\wedge \of x^{c_{d-q}}
\ee
When $p=q$ this simplifies.
\be
\alpha\wedge \star \beta = \alpha_{a_1\dots a_p}\beta^{a_1\dots a_p}\frac{\sqrt{|g|}}{p!}\of x^{1}\wedge\dots\wedge \of x^{d} ~.
\ee
The interior product $\iota_X$ is defined as 
\be
\iota_X \alpha = \frac1{(p-1)!}X^a\alpha_{ab_1\dots b_{p-1}} \of x^{b_1}\wedge \dots\wedge \of x^{b_{p-1}} ~,
\ee

\subsection{Iyer-Wald formalism} 
We start with writing the Lagrangian for a $d$-dimensional gravitational theory with arbitrary matter fields $\phi$ as a $d$-form ${\bf L}$. Varying with respect to a matter field results in 
\be
\delta {\bf L} = {\bf E}(\delta\phi) + \of{\bf \Theta}(\delta\phi) ~.
\ee
Here ${\bf E}$ collectively denotes the equations of motions and $\bf \Theta$ is the so-called symplectic potential. An anti-symmetric variation of the symplectic potential yields the symplectic current
\be
\omega(\delta_1\phi,\delta_2\phi) = \delta_1{\bf \Theta}(\delta_2\phi) - \delta_2{\bf \Theta}(\delta_1\phi) ~.
\ee
Now consider an infinitesimal diffeomorphism labeled by a vector field $\xi$, which acts as $\delta_\xi\phi = {\cal L}_\xi\phi$. Integrating the symplectic current over a Cauchy surface $\Sigma$ gives the symplectic form, which with foresight we will write as the variation of an Hamiltonian that generates the flow of $\xi$.
\be \label{eq:hamiltonian}
\delta H_\xi = \int_\Sigma \omega(\delta\phi,{\cal L}_\xi\phi) ~.
\ee
For any $\xi$ we can construct a Noether current
\be \label{eq:Noethercurrent}
{\bf J}_\xi = {\bf\Theta}({\cal L}_\xi\phi) - \iota_\xi {\bf L} ,
\ee
which is conserved on-shell.
\be
\of{\bf J}_\xi = -{\bf E}(\cal L_\xi) ~.
\ee
The fact that ${\bf J}_\xi$ is closed (on-shell) and only depends linearly on $\xi$ implies that we can write it as\footnote{See page 21 of \cite{Compere:2018aar} for the proof.}
\be
{\bf J}_\xi = -{\bf E}({\cal L_\xi\phi}) + \of{\bf Q}_\xi({\cal L}_\xi\phi) ~,
\ee
where ${\bf Q}_\xi$ is the Noether charge. To extract conserved quantities from the Noether current, we consider a variation
\be
\delta {\bf J}_\xi = \of\iota_\xi {\bf \Theta}(\delta\phi) - \iota_\xi {\bf E}(\delta\phi) + \omega(\delta\phi,{\cal L}_\xi\phi) ~.
\ee
On-shell, the symplectic current can be written as
\be
\omega(\delta\phi,{\cal L}_\xi\phi) = \of(\delta{\bf Q}_\xi) - \of(\iota_\xi{\bf \Theta}(\delta\phi)) ~.
\ee
and the variation of the Hamiltonian is
\be \label{eq:varHamil}
\delta {\bf H}_\xi = \delta{\bf Q}_\xi - \iota_\xi{\bf \Theta}(\delta\phi) ~.
\ee
When $\xi$ is a symmetry, i.e. ${\cal L}_\xi\phi=0$, the Hamiltonian is conserved
\be 
\of\delta {\bf H}_\xi = \omega(\delta\phi,{\cal L}_\xi\phi) = 0 ~.
\ee
We therefore see that \eqref{eq:hamiltonian} indeed gives the conserved quantities.

\subsection{Einstein-Maxwell gravity}
Let us now restrict to Einstein-Maxwell gravity. The Lagrangian is given by
\be
{\bf L} = \frac1{2\kappa^2}\left(R-2\Lambda\right)\epsilon -\frac12F\wedge \star F~.
\ee
Here $\kappa^2=8\pi G_d$, with $G_d$ the $d$-dimensional Newton constant. We can now perform variations with respect to the metric $\delta g_{ab}=h_{ab}$ and the gauge field $\delta A_a$. Also, in addition to diffeomorphisms we can perform gauge transformations on the gauge field $\delta_\lambda A=\of\lambda$. Varying with respect to the metric we find
\be \label{eq:symplecticg}
{\bf E}_g(h) = -E^{ab}h_{ab} \epsilon ~, \quad {\bf \Theta}_g(h) = \iota_X\epsilon \nn ~,
\ee
with
\begin{align}
E_g^{ab} &= \frac1{2\kappa^2}\left(R^{ab}-\frac12g^{ab}(R-2\Lambda)\right) +\frac18 g^{ab}F_{cd}F^{cd}- \frac12 F^{ac}F^b_{\,\,\,c} ~,\\
X^a &= \frac1{2\kappa^2}\left(\nabla_b h^{ab} - \nabla^ah^b_{\,\,\,b}\right) ~. \nn
\end{align}
Varying with respect to the gauge field we find
\begin{align}\label{eq:symplecticA}
{\bf E}_A(\delta A) =  -\delta A\wedge \of{\star F} ~, \quad {\bf\Theta}_A(\delta A) = -\delta A\wedge \star F ~.
\end{align}
From \eqref{eq:Noethercurrent} we can now construct the Noether current. The Noether current is now given by
\be
{\bf J} = {\bf\Theta}_g({\cal L}_\xi g) + {\bf\Theta}_A({\cal L}_\xi A) + {\bf\Theta}_A(\of\lambda) - \iota_\xi {\bf L} .
\ee
The expressions for the Lie derivatives are ${\cal L}_\xi g_{ab} = 2\nabla_{(a}\xi_{b)}$ and ${\cal L}_\xi A_a = \xi^bF_{ba} + \partial_a(\xi^bA_b)$. Using these we find
\begin{align}
{\bf \Theta}_A({\cal L}_\xi A) &= -\left(\iota_\xi F + \of(\iota_\xi A) \right)\wedge \star F ~, \\
{\bf\Theta}_A(\of \lambda) &= -\of\lambda\wedge \star F ~, \nn \\
{\bf\Theta}_g({\cal L}_\xi g) &= \frac{1}{2\kappa^2}\left(2\nabla_b\nabla^{(b}\xi^{a)}-2\nabla^a\nabla_b\xi^b\right)\frac{\sqrt{|g|}}{(d-1)!}\epsilon_{ab_1\dots b_{d-1}}\of x^{b_1}\wedge\dots\wedge \of x^{b_{d-1}} ~. \nn
\end{align}
After some algebra, the Noether current can be written as\footnote{The superscript $\flat$ denotes the one-form dual $\xi^\flat = g_{ab}\xi^b\of x^a$ and $E_g\cdot\xi=E_{ab}\xi^b\of x^a$.}
\begin{align}
{\bf J} &= 2 \star (E_g\cdot\xi) -\frac1{2\kappa^2}\of{\star \of\xi^\flat} +\left(\iota_\xi A+\lambda\right)\of{\star F} -\of\left[\left(\iota_\xi A+\lambda\right){\star F}\right]  \nn~.
\end{align}
We see that the Noether charges for $\xi$ and $\lambda$ are given by.
\begin{align} \label{eq:charges}
{\bf Q}_\xi &= -\frac1{2\kappa^2}{\star \of\xi^\flat} ~, \\
{\bf Q}_\lambda &= -(\iota_\xi A+\lambda)\star F  \nn ~.
\end{align}
Using \eqref{eq:varHamil}, the variation of the Hamiltonian is given by
\begin{align}\label{eq:varHamil2}
\delta{\bf H}_\xi &= - \frac1{2\kappa^2}\left(\delta(\star\of \xi^\flat) +\iota_\xi\iota_X \epsilon\right) ~,  \\
\delta{\bf H}_\lambda &=  -(\iota_\xi A+\lambda)\star \delta F \nn ~.
\end{align}
Taking an exterior derivative of the variation of the Hamiltonian we obtain
\be \label{eq:appconservation}
\of \delta{\bf H} = -2\delta\star(E_g\cdot\xi) -(\iota_\xi A+\lambda)\,\of{\star \delta F}  ~,
\ee
which vanishes on-shell.

\section{Five-dimensional black string with higher-derivative corrections}\label{app:explicit5Dstring}
In this appendix we provide some of the details of the $\alpha$-corrected black string in five dimensions. This brute-force solving of the equations of motion reproduces the thermodynamic and $c$-extremization arguments as presented in the main text.

Begin by writing the five-dimensional action with higher-derivative terms as
\begin{align}
I = \frac1{16\pi G_5}\int \of^5x\,\sqrt{-g}\,&\Big(R-\frac34F_{MN}F^{MN} + \alpha_1Q^2F_{MN}F^{MN}F_{OP}F^{OP} \\
& {}+ \alpha_2 Q^2 F_{MN}F_{OP}W^{MNOP} + \alpha_3 Q^2 E_5 \Big) ~. \nn
\end{align}
where $E_5=R_{\mu\nu\rho\sigma}R^{\mu\nu\rho\sigma} - 4R_{\mu\nu}R^{\mu\nu} + R^2$ is the Euler density. As in Sec.~\ref{sec:hdcorrectedblackstring} we normalize the Wilson coefficients with $Q$ so that the $\alpha_i$ are dimensionless. Take for an ansatz
\begin{equation}
\begin{aligned}
    \d{s^2} &= H^{-1}\big({-f}\,\d{t^2} + h\,\d{x^2}\big) + H^2\big(f^{-1}\,\d{r^2} + r^2\,\d{\Omega_2^2}\big) \,,\\
    F &= \sqrt{q(q+r_0)} \,\sin{\theta}\,\d{\theta}\wedge\d{\phi} \,.
\end{aligned}
\end{equation}
The leading-order ($\alpha_i=0$) solution has
\begin{equation}
    H(r) = 1 + \frac{q}{r} \,, \qquad h(r) = 1 \,, \qquad f(r) = 1 - \frac{r_0}{r} \,.
\end{equation}
Before turning to the $\mathcal{O}(\alpha)$ corrections to these functions, let us discuss how these functions appear in the near-horizon limit and 4D reduction. In boosting the string along the $x$-direction we make the replacements
\begin{equation}
\begin{aligned}
    t &\to \cosh{\delta_0}\,t + \sinh{\delta_0}\,x \,,\\
    x &\to \sinh{\delta_0}\,t + \cosh{\delta_0}\,x \,,
\end{aligned}
\end{equation}
which brings the metric to the form
\begin{align}
    \d{s^2} &= H^{-1}\Big[ {-\d{t^2}} + \d{x^2}\\
    &\qquad + (1-f)\big(\cosh{\delta_0}\,\d{t} + \sinh{\delta_0}\,\d{x}\big)^2 + (h-1)\big(\sinh{\delta_0}\,\d{t} + \cosh{\delta_0}\,\d{x}\big)^2\Big] \notag\\
    &\qquad\qquad + H^2\big(f^{-1}\,\d{r^2} + r^2\,\d{\Omega_2^2}\big) \notag\\
    \label{eq:5DmetricApp}
    &= H^{-1}\left[ {-H_0^{-1}fh}\,\d{t^2} + H_0\big( \d{x} + H_0^{-1}(h-f)\sinh{\delta_0}\cosh{\delta_0}\,\d{t} \big)^2 \right]\\
    &\qquad\qquad +H^2\big(f^{-1}\,\d{r^2} + r^2\,\d{\Omega_2^2}\big) \,, \notag
\end{align}
where we have introduced
\begin{equation}
    H_0(r) = h(r)\cosh^2{\delta_0} - f(r)\sinh^2{\delta_0} \,.
\end{equation}

\paragraph{Reduction to four dimensions}
From equation~\ref{eq:5DmetricApp} we may read off the KK photon profile,
\begin{equation}\label{eq:A0tApp}
    (A_0)_t = H_0^{-1}(h-f)\sinh{\delta_0}\cosh{\delta_0} = \frac{\sqrt{q_0(q_0+r_0)}}{r+q_0} + \mathcal{O}(\alpha) \,,
\end{equation}
where $q_0=r_0\sinh^2{\delta_0}$. After reducing and going to Einstein frame the metric reads
\begin{equation}
    \d{s^2}\big|_{\text{4D}} = -(H^3H_0)^{-1/2}fh\,\d{t^2} + (H^3H_0)^{1/2}\big(f^{-1}\,\d{r^2} + r^2\,\d{\Omega_2^2}\big) \,.
\end{equation}

\paragraph{Near-horizon limit}
The near-horizon geometry is found by taking $q\gg r,r_0$. In this limit the metric splits into a (locally) AdS$_3$ space and constant-radius $S^2$.

\paragraph{Boundary conditions}
In solving for the corrected five-dimensional solution we should keep the asymptotic form of the solution fixed, namely $H,h,f=1 + \mathcal{O}(\frac{1}{r})$. To work with fixed charges, we should also impose that the $\frac{1}{r}$ term in Equation~\ref{eq:A0tApp} is uncorrected. It is also convenient to choose coordinates (equivalently, integration constants) so that the outer horizon remains at $r=r_0$ and extremality is still $r_0\to0$. With these choices the $\alpha$-corrected solutions are uniquely determined. The full expressions for the corrected $H,h,f$ are quite cumbersome, so we present here only their form in some relevant limits, as needed.

\paragraph{Corrected reduction to four dimensions}
In the asymptotic region, $r\gg q,r_0$, the corrected solution reads
\begin{align}
    H(r) &= 1 + \frac{q}{r} - \frac{Q^2}{(2q+r_0)r}G(\alpha_i;\tfrac{r_0}{q}) - \frac{Q^2}{4r^2}G(\alpha_i;\tfrac{r_0}{q}) + \mathcal{O}\Big(\frac{1}{r^3}\Big) \,,\\
    h(r) &= 1 + \frac{Q^2}{4r^2}G(\alpha_i;\tfrac{r_0}{q}) + \mathcal{O}\Big(\frac{1}{r^3}\Big) \,,\\
    f(r) &= 1 - \frac{r_0}{r} + \frac{Q^2}{4r^2}G(\alpha_i;\tfrac{r_0}{q}) + \mathcal{O}\Big(\frac{1}{r^3}\Big) \,.
\end{align}
The function $G(\alpha_i;z)$ has the following small-$z$ limit, relevent for the extremal limit $r_0\to0$:
\begin{equation}
    G(\alpha_i;z) = \frac{8\alpha_1+7\alpha_2+6\alpha_3}{15} + \mathcal{O}(z^{3/2}) \,.
\end{equation}
Choosing $q_0=q$ for simplicity, the ADM mass can be read off from the $\frac{1}{r}$ coefficent of $(H^3H_0)^{1/2}f^{-1}$:
\begin{equation}
    2G_4M_4 = 2q + r_0 - \frac{3}{4}(q+r_0)\,G(\alpha_i;\tfrac{r_0}{q}) \,.
\end{equation}
Taking $r_0\to0$, we find that the four-dimensional extremality bound is corrected to
\begin{equation}
    \frac{Q}{G_4M_4} \leq 1 + \frac{3}{8}G(\alpha_i;0) = 1 + \frac{8\alpha_1+7\alpha_2+6\alpha_3}{40} \,.
\end{equation}

\paragraph{Corrected near-horizon limit}
With $q\gg r\gg r_0$, the corrected solution behaves as\footnote{The inequality $r\gg r_0$ may seem odd for a near-horizon limit, but we are ultimately interested in the extremal limit, $r_0\to0$, so that $r\gg r_0$ is satisfied for any finite $r$. Note also that the singularity is at $r=-q$.}
\begin{equation}
    H(r) = \left(1 + \frac{Q}{r}\right) - \frac{Q(8\alpha_1+3\alpha_2-2\alpha_3)}{4r} + \cdots \,,
\end{equation}
so that we may read off that the extremal $S^2$ radius has been corrected to
\begin{equation}
    \ell_{S^2} = Q\left(1 - \frac{8\alpha_1+3\alpha_2-2\alpha_3}{4}\right) \,.
\end{equation}
The corrected AdS$_3$ length is most easily found by looking at the Ricci scalar, since $R_{\text{BTZ}\times S^2} = R_\text{BTZ} + R_{S^2}$ in the near-horizon region. On the corrected solution we find
\begin{equation}
\begin{aligned}
    R_5 &= \frac{q(q+r_0)}{2(q+r)^4} + \mathcal{O}(\alpha) = \frac{1}{2Q^2} + \frac{8\alpha_1 + 2\alpha_2+\alpha_3}{Q^2} + \cdots = \left( -\frac{6}{\ell^2} \right) + \left( \frac{2}{\ell_{S^2}^2} \right) + \cdots \,,
\end{aligned}
\end{equation}
where $\ell_{S^2}$ from above can be used to isolate the (corrected) AdS$_3$ length,
\begin{equation}
    \ell = 2Q\left(1 - \frac{2(\alpha_2-3\alpha_3)}{3}\right) \,.
\end{equation}
As expected, the near-horizon AdS$_3$ and $S^2$ lengths found by brute force agree with the (much simpler) $c$-extremization.

\bibliographystyle{JHEP}
\bibliography{refs}

\end{document}